# Simulation of hydrogen permeation through pure iron for trapping and surface phenomena characterisation

A. Díaz[1], A. Zafra[2], E. Martínez-Pañeda[3], J.M. Alegre[1], J. Belzunce[2], I.I. Cuesta[1]

[1]University of Burgos, Escuela Politécnica Superior. Avenida Cantabria s/n, 09006, Burgos, Spain

[2]University of Oviedo. Campus Universitario, EPI, 33203, Gijón, Spain

[3]Imperial College London, Department of Civil and Environmental Engineering. London SW7 2AZ, UK

**Abstract**

There is a need for numerical models capable of predicting local accumulation of hydrogen near stress concentrators and crack tips to prevent and mitigate hydrogen assisted fracture in steels. The experimental characterisation of trapping parameters in metals, which is required for an accurate simulation of hydrogen transport, is usually performed through the electropermeation test. In order to study grain size influence and grain boundary trapping during permeation, two modelling approaches are explored; a 1D Finite Element model including trap density and binding energy as input parameters and a polycrystalline model based on the assignment of a lower diffusivity and solubility to the grain boundaries. Samples of pure iron after two different heat treatments – 950ºC for 40 minutes and 1100ºC for 5 minutes – are tested applying three consecutive rising permeation steps and three decaying steps. Experimental results show that the finer grain microstructure promotes a diffusion delay due to grain boundary trapping. The usual methodology for the determination of trap densities and binding energies is revisited in which the limiting diluted and saturated cases are considered. To this purpose, apparent diffusivities are fitted including also the influence of boundary conditions and comparing results provided by the constant concentration with the constant flux assumption. Grain boundaries are characterised for pure iron with a binding energy between 37.8 and 39.9 kJ/mol and a low trap density but it is numerically demonstrated that saturated or diluted assumptions are not always verified, and a univocal determination of trapping parameters requires a broader range of charging conditions for permeation. The relationship between surface parameters, i.e. charging current, recombination current and surface concentrations, is also studied showing that trapping phenomena are stronger during the diluted steps and that recombination currents are much higher than the steady state obtained flux.

## 1. Introduction

Structural integrity of metallic components and structures can be drastically affected by the effects of hydrogen due to the degradation mechanisms operating when hydrogen diffuses through the bulk material. The most challenging phenomenon is hydrogen embrittlement, also named as hydrogen assisted cracking, in which a toughness reduction and an increase in crack growth rate is cause by atomic diffusing hydrogen without the presence of other processes such as $H_2$ or methane combination, blistering, hydride formation, etc. The underlying micro-mechanisms of hydrogen embrittlement are not completely understood but the process appears to be driven by local hydrogen concentration. It has been empirically proved that hydrogen accumulation near the fracture process zone is the triggering process for cracking [1]. Therefore, many efforts have been dedicated to model hydrogen transport near stress concentrators and crack tips [2–6], including the delaying effects of metal defects such as dislocations, grain boundaries, inclusions or vacancies [7]. The apparent lower diffusivity caused by defects is explained by the lower potential energy of hydrogen in these "traps", so hydrogen atoms are retained because the hop probability is low in comparison to the motion in the ideal crystal lattice; whether trapping promotes fracture reduction or mitigates embrittlement depends on the nature of defects [8,9].

Characterisation of trapping sites for hydrogen is usually carried out using electro-permeation (EP) or thermal desorption spectroscopy (TDS). In both cases, hydrogen is not locally resolved [10], and trapping features are determined by fitting output fluxes or desorption rates, respectively, to numerical solutions of the associated mass diffusion problem. Electro-permeation is a very common technique due to its simplicity and low cost. It is based on a two-cell setup, as proposed by Devanathan and Stachurski [11], and has been standardised. However, the usual permeation methodology has some limitations hindering complete trap characterisation; the procedure relies on the numerical fitting of permeation transients to analytical expressions that assume a constant diffusivity coefficient, denominated as apparent diffusivity in this work, $D_{app}$. This is a phenomenological descriptor that, even though it is useful for a first trapping assessment, has been demonstrated to depend on concentration and charging conditions, so it cannot be univocally used to characterise material traps. Additionally, two-level numerical models [2,3,5] are not easy to adapt because the determination of trap densities and binding energy requires some assumptions [12]. On the other hand, hydrogen entry is sometimes overlooked. Output fluxes are usually normalised, and the implications of steady state values are not assessed or just used to calculate apparent concentration. Generalised boundary conditions from electrochemical theory can shed light into this problem [13]. Since the multi-trapping effects produce many complex interactions, in the present work pure iron is analysed after two different heat treatments to obtain different grain sizes and thus different fractions of grain boundary surfaces.

## 2. Experimental procedure

### 2.1. Material

50x50 mm$^2$ sheet samples of pure iron (99.5 %) with a thickness of 1 mm underwent two different annealing treatments in order to obtain homogeneous microstructures with different grain sizes. Consequently, one sample was maintained at 925ºC for 40 min followed by furnace cooling and another sample was kept at 1100ºC for 5 min also followed by furnace cooling. In order to analyse permeation results for both samples, heat treatment temperatures for the different grain sizes are identified as $T_g$ throughout the paper.

After carefully cutting the specimen (to avoid microstructural alteration), both samples were metallographically prepared (ground and polished onto synthetic cloths with 6 and 1 μm diamond pastes) and etched with Nital 2%. Their microstructures were observed using an optical microscope (Nikon Epiphot 300) and a scanning electron microscope (SEM JEOL-JSM5600) under an acceleration voltage of 20 kV, as shown in Figure 1. Additionally, the average grain size was determined in both cases following the ASTM E112-13 standard [14].

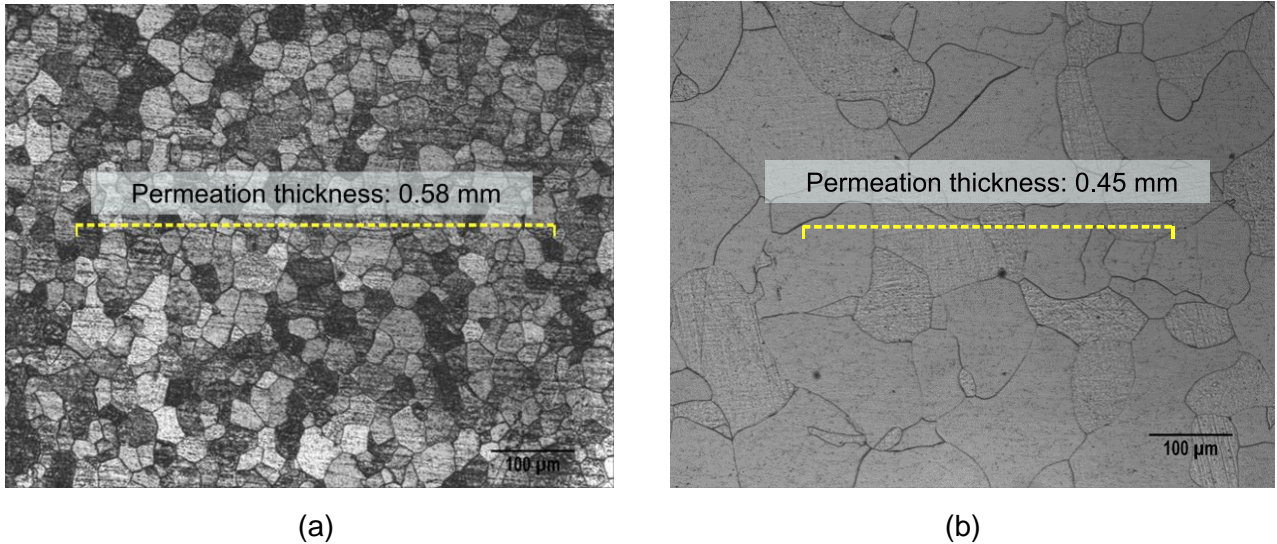

(a)                                         (b)

Figure 1. Microstructure of the samples observed in the optical microscope for samples maintained at: (a) 925ºC for 40 minutes and, (b) 1100ºC for 5 minutes. The permeation thickness scale is sketched on the microstructure in order to estimate the number of grains involved in permeation.

### *2.2. Permeation setup*

The hydrogen transport and trapping behaviour of pure iron with different grain sizes was characterized by means of electrochemical permeation tests. Flat specimens measuring 20x20 mm were machined and ground up to 1200 grit SiC paper until attaining a final sample thickness of 0.6-0.45 mm. A circular exposed area of 1.25 cm$^2$ was always used.

The permeation tests were performed in a double electrolytic cell based on the one developed by Devanathan and Stachurski [11,15], as shown in Figure 2. With an approximate volume of 300 ml, both cells satisfy the ASTM G148-97(2018) [16] recommendation of a solution volume-to-surface area ratio greater than 20 ml/cm$^2$.

Both cells are separated by the specimen, which is the working electrode (WE) in each cell. The cathodic cell, where hydrogen is cathodically generated and adsorbed on the surface of the metal via the application of a cathodic current, was filled with an acid solution (pH≈1) composed of 1M $H_2SO_4$ and 0.25g/l $As_2O_3$ to mitigate hydrogen recombination reactions. The other side of the specimen, the anodic cell, where hydrogen oxidation occurs, was filled with a basic solution (pH≈12.5) of 0.1M NaOH. Thin platinum plates with a total surface area of 1 cm$^2$ (similar to the specimen's permeated area) were used as counter electrodes (CE). A reference silver-silver chloride electrode (Ag/AgCl, RE) with a Luggin capillary was employed in the anodic cell and the equipment used for data acquisition was a pocketSTAT Ivium potentiostat with a current operation range of ±10 mA. All tests were performed at room temperature.

Before starting the tests, it is necessary to decrease the background current density in the anodic cell to a steady-state value below 0.1 µA/cm$^2$ (which must be subtracted from the measured oxidation current prior to data analysis). To this end, an homogeneous palladium coating (around 1-2 µm thick measured by SEM) was electrodeposited on the anodic side of the sample from a commercial palladium bath containing 2 g/l Pd, applying a current density of 3 mA/cm$^2$ for 5 min. Hydrogen oxidation is thereby enhanced in the anodic cell, ensuring a virtually zero hydrogen concentration on the exit side of the specimen. In fact, there is general consensus as to the importance of using palladium coatings on the detection side of ferrous samples so that the permeation results may be reliably exploited, in order to ensure the oxidation of hydrogen atoms on palladium-coated surfaces under most charging conditions [17]. Moreover, the possibility of having introduced hydrogen in the sample during the process of Pd electrodeposition was discarded, as different hydrogen measurements were performed on the Pd-coated samples obtaining values below 0.1 ppm in all cases.

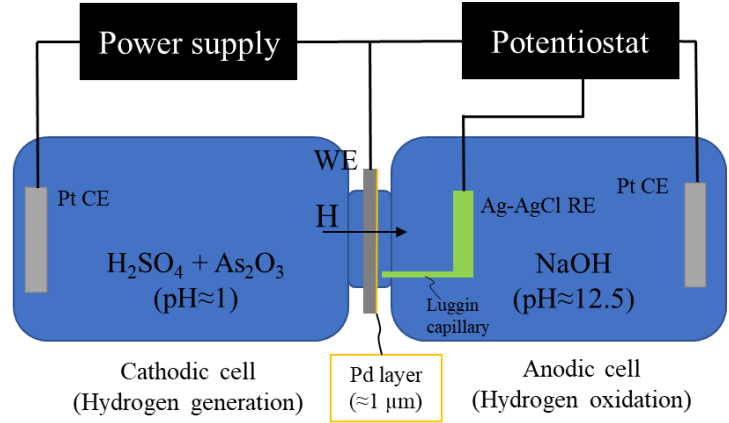

Figure 2. Scheme of the modified D-S double-cell employed in the hydrogen permeation experiments.

A fixed current density is imposed in the entry side (galvanostatic charging condition). Three rising steps and three consecutive decaying steps were used. Three partial build-up permeation transients were applied by sequentially increasing the cathodic current density (0.5 + 0.5 + 1 mA/cm$^2$) up to a final cathodic current density of 2 mA/cm$^2$. This was followed by three consecutive decay transients (under analogous cathodic current density drops). The choice of three rising – three decaying steps follows [18], and aims at covering different trapping regimes without many steps for the sake of clarity.

## 3. Numerical methodology

The numerical procedure aims at determining characteristic trapping parameters of pure iron with two different grain sizes and to identify the limitations of the common analysis methods of electrochemical permeation transients. The output magnitude that is being registered during experimental permeation is the exit current measured in the oxidation cell. This current is divided by the exposed area to find the current density, $i_p$, in µA/cm$^2$. It is assumed that the flux of egressed hydrogen atoms is proportional to this current density, i.e. $j(t) = i_p(t)/F$, where $F$ is the Faraday's constant, so the flux takes units of [mol·m$^{-2}$·s$^{-1}$].

In a first approximation, steps are separated and scaled so they are analysed independently. The normalisation is performed considering the initial flux, $j_0$ and the steady state value $j_{ss}$ for each individual step. Then, the obtained transient is fitted considering analytical solutions of 1D diffusion in an exponential series form, as discussed in Section 3.2. The fitted apparent diffusivity $D_{app}$ can be used to indirectly determine a concentration in the entry surface, which is named as $C_{app}$. Previously, a two-level modelling approach is presented in Section 3.1, with the aim of establishing a relationship between apparent diffusivity and two important trapping features: density, $N_T$, and binding energy, $E_b$. However, the validity of the required assumptions, i.e. that traps are saturated or diluted [12,19], must be discussed for each specific case. To facilitate this discussion, a trapping regime identification is proposed following the mapping framework from different authors [12,20,21].

Finally, a finite element framework is used to simulate hydrogen permeation considering both 1D two-level model that includes trapping effects and the 2D polycrystal model; the latter is presented in Section 3.5. and the method for determining grain boundary segregation $s_{gb}$ and diffusivity $D_{gb}$ is discussed. Hydrogen entry is taken into account by defining appropriate boundary conditions; the constant concentration assumption, $C_L^0$, is discussed in contrast to a constant flux that depends on charging ($i_c$) and recombination ($i_r$) currents at the entry side. The numerical permeation transients, that have been informed with the fitted trapping and charging parameters, are thus compared to the experimental curves. The complete flowchart is shown in Figure 3.

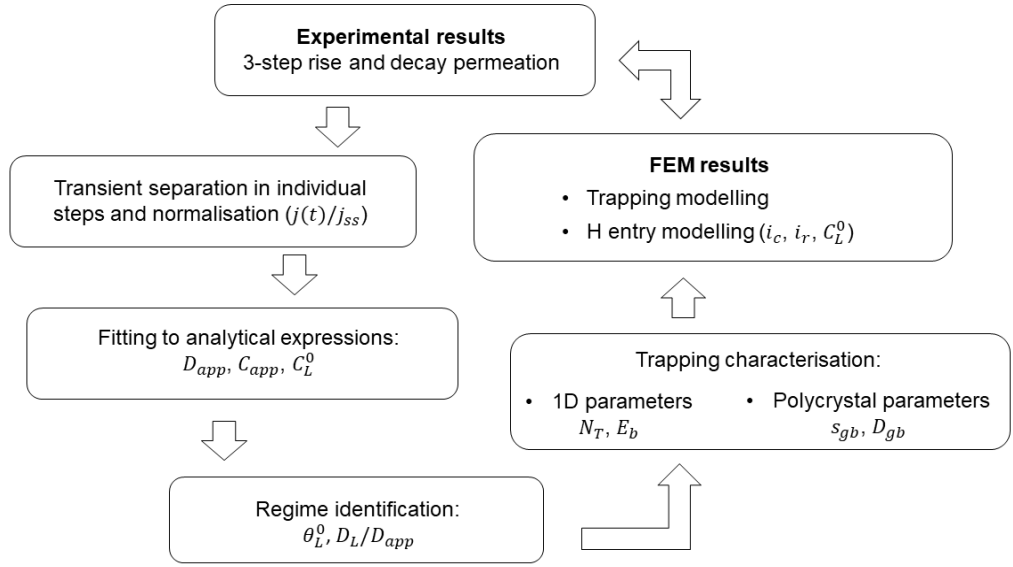

Figure 3. Flowchart of the permeation analysis with numerical models informed by fitted parameters and the final comparison between simulated and experimental transients.

### 3.1. Continuum 1D model

Permeation is numerically solved in Comsol Multiphysics where the associated heat equation, i.e. a parabolic equation with flux, source and capacity terms, is implemented. For unidimensional diffusion in $x$ direction and for a constant lattice diffusivity $D_L$, the mass balance can be written as:

$$\frac{\partial C_L}{\partial t} + \frac{\partial C_T}{\partial t} = -\frac{\partial j}{\partial x} \qquad (1)$$

where $C_L$ and $C_T$ are hydrogen concentrations in lattice and trapping sites, respectively. Hydrogen flux $j$ follows the Fick's law:

$$j = -D_L \frac{\partial C_L}{\partial x} \qquad (2)$$

Considering that both the density of lattice sites and trapping sites, $N_L$ and $N_T$ respectively, remain constant and do not vary over time, the governing equation can be expressed in terms of lattice and trapping occupancies that are respectively defined as $\theta_L = C_L/N_L$ and $\theta_T = C_T/N_T$:

$$\frac{\partial \theta_L}{\partial t} + \frac{N_T}{N_L}\frac{\partial \theta_T}{\partial t} = \frac{\partial}{\partial x}\left(D_L \frac{\partial \theta_L}{\partial x}\right) \tag{3}$$

The source term can be transformed to a capacity term:

$$\left(\frac{N_T}{N_L}\frac{\partial \theta_T}{\partial \theta_L} + 1\right)\frac{\partial \theta_L}{\partial t} = D_L \frac{\partial^2 \theta_L}{\partial x^2} \tag{4}$$

A normalisation is followed for the sake of generalisation. Non-dimensional time and distance are scaled considering lattice diffusivity and the specimen thickness:

$$\bar{t} = tD_L/L^2 \tag{5}$$

$$\bar{x} = x/L \tag{6}$$

The scaling defined in Equations (5) and (6) has already been performed by other authors for hydrogen transport models [19,21–23]. The non-dimensional governing equation is then expressed as:

$$\left(\frac{N_T}{N_L}\frac{\partial \theta_T}{\partial \theta_L} + 1\right)\frac{\partial \theta_L}{\partial \bar{t}} = \frac{\partial^2 \theta_L}{\partial \bar{x}^2} \tag{7}$$

And the term $\partial \theta_T/\partial \theta_L$ is found by assuming thermodynamic equilibrium between trapping and lattice sites:

$$\frac{\theta_T}{1-\theta_T} = \frac{\theta_L}{1-\theta_L}K \tag{8}$$

where $K = \exp(E_b/RT)$ and the trapping binding energy $E_b$ is considered positive. Assuming a low lattice occupancy, $\theta_L \ll 1$:

$$\theta_T = \frac{\theta_L K}{1+\theta_L K} \tag{9}$$

Deriving $\partial \theta_T/\partial \theta_L$ and substituting into (7), the governing non-dimensional equation is:

$$\left(\frac{N_T K}{N_L(1+\theta_L K)^2} + 1\right)\frac{\partial \theta_L}{\partial \bar{t}} = \frac{\partial^2 \theta_L}{\partial \bar{x}^2} \tag{10}$$

At this point, it must be highlighted that the capacity term for the parabolic heat equation is only treatable when the expression $\partial \theta_T/\partial \theta_L$ can be defined; for generalised kinetic approaches, i.e. following McNabb and Foster's formulation [19], or even a more general framework [24,25], the source term including $\partial \theta_T/\partial t$ mut be considered. For the assumed thermodynamic equilibrium, an effective diffusivity can be defined as:

$$D_{eff} = \frac{D_L}{1 + \frac{N_T}{N_L}\frac{\theta_T}{\theta_L}(1-\theta_T)} \tag{11}$$

When traps are completely filled, i.e. $\theta_T \approx 1$, $D_{eff} = D_L$. On the other hand, when the trap occupancy is low, $\theta_T \ll 1$, the effective diffusivity can be expressed as:

$$D_{eff} = \frac{D_L}{1 + \frac{N_T}{N_L}K} \qquad (12)$$

which is a magnitude independent of local concentration and only depends on material parameters and temperature. Even though the equation (10) is already non-dimensional, occupancy values $\theta_L$ are usually very low. Thus, the lattice occupancy can also be scaled, $\bar{\theta}_L = \theta_L/\theta_L^0$, when the constant concentration assumption is adopted, i.e. a fixed $\theta_L^0 = C_L^0/N_L$ is imposed in the entry surface. The implemented equation in the Finite Element code is then:

$$\left(\frac{N_T K}{N_L(1 + \bar{\theta}_L \theta_L^0 K)^2} + 1\right)\frac{\partial \bar{\theta}_L}{\partial \bar{t}} = \frac{\partial^2 \bar{\theta}_L}{\partial \bar{x}^2} \qquad (13)$$

The lattice occupancy in the entry boundary, $\theta_L^0$, must not be confused with an initial concentration. For electro-permeation simulations, $\theta_L(t=0) = 0$. This normalisation is possible for both constant concentration (CC) and constant flux (CF) models that will be described in the following sections; for the former, $\theta_L^0$ corresponds to the $\theta_L$ value prescribed at the boundary while for the constant flux approcach, $\theta_L^0$ represents the lattice occupancy that will be reached at the steady state and it can be determined by assuming that $j_{in} = j_{ss}$ and thus $\theta_L^0 = j_{ss}L/(N_L D_L)$. Therefore, non-dimensional flux is also redefined as:

$$\bar{j} = -\frac{\partial \bar{\theta}_L}{\partial \bar{x}} \qquad (14)$$

The flux scaling within the FE framework must not be confused with the normalisation considering the steady state flux that is used in the following section in order to fit experimental transients to analytical expressions.

### 3.2. Analytical fitting

Permeation transients that have been experimentally measured are usually fitted to analytical expressions in order to determine a diffusion coefficient. Throughout the present work, this phenomenological parameter is named as apparent diffusivity, $D_{app}$, with the aim of avoiding confusion with other local diffusivities. The analytical approach assumes that the time-dependent solution only depends on $D_{app}$ and on the thickness of the specimen $L$; thus, the output flux follows a function $f(D_{app}, L, t)$. It is also possible to define a non-dimensional time, $\tau = tD_{app}/L^2$, so a single-variable function governs the problem, $f(\tau)$. Non-dimensional time $\tau$, which depends on the fitted apparent diffusivity, must not be confused with the scaled time $\bar{t}$, which is used for the FE implementation. All permeation steps are also normalised considering the flux at the beginning of the step $j_0$ and the steady state flux $j_{ss}$. For rising steps ($j_0 \leq j(t) \leq j_{ss}$):

$$\frac{j(t) - j_0}{j_{ss} - j_0} = f(D_{app}, L, t) \qquad (15)$$

For decaying steps ($j_{ss} \leq j(t) \leq j_0$):

$$\frac{j(t) - j_{ss}}{j_0 - j_{ss}} = 1 - f(D_{app}, L, t) \tag{16}$$

It must be noted that time $t$ for permeation fitting represents the individual time for each step, i.e. $t = t_{total} - t_0$. The function to be fitted is the analytical solution of Fick's laws for 1D permeation and it is defined using a series expansion form. The solution depends on the problem boundary conditions; for a constant concentration on the entry side [26–28]:

$$f(D_{app}, L, t) = 1 + 2 \sum_{n=1}^{\infty} (-1)^n \exp\left(-\frac{n^2 \pi^2 D_{app} t}{L^2}\right) \tag{17}$$

Fitting to this expression is recommended in both ASTM G148-97(2018) and ISO 17081:2014 Standards for electrochemical permeation. An alternative that does not require fitting algorithms applies the relationship between a permeation time $t_i$ and diffusivity:

$$D_{app} = \frac{L^2}{M t_i} \tag{18}$$

The most used permeation times are the breakthrough time $t_b = t_{0.10}$, defined as the time when the 10% of the maximum flux is reached, and the lag time $t_{lag} = t_{0.63}$, that corresponds to the 63% of the transient steady state. The constant takes a value $M = 15.3$ for $t_{0.10}$ and $M = 6$ for $t_{0.63}$ [20]. It must be taken into account that these $M$ values, since they are derived from equation (17), assume a constant concentration at the entry side. An alternative analytical solution for constant concentration is expressed as [29,30]:

$$f(D_{app}, L, t) = \frac{2L}{\sqrt{\pi D_{app} t}} \sum_{n=0}^{\infty} \exp\left(-\frac{(2n+1)^2 L^2}{4 D_{app} t}\right) \tag{19}$$

The second series, i.e. equation (19), converges more rapidly for small times [27]. On the contrary, the analytic solution for a constant flux in the entry side reads [27]:

$$f(D_{app}, L, t) = 1 - \frac{4}{\pi} \sum_{n=0}^{\infty} \frac{(-1)^n}{2n+1} \exp\left(-\frac{(2n+1)^2 \pi^2 D_{app} t}{4 L^2}\right) \tag{20}$$

or, alternatively, for small times [27]:

$$f(D_{app}, L, t) = 2 \sum_{n=0}^{\infty} (-1)^n \operatorname{erfc}\left(-\frac{(2n+1) L}{\sqrt{D_{app} t}}\right) \tag{21}$$

In Section 4.1, the experimental rise and decaying steps for hydrogen permeation through pure iron are analysed considering these three approaches described above: (i) constant concentration (CC) fitting for long times, i.e. equation (17) (ii) diffusivities from the $t_{0.63}$ method, that assumes also CC but does not require fitting and (iii) constant flux (CF) fitting for long times, i.e. equation (20).

### 3.3. Determination of trapping parameters from $D_{app}$ and $j_{ss}$

Trapping effects can also be fitted by considering a relationship between trap density, permeation times and charging conditions. The simplified expression that has been derived by McNabb and Foster [19] in their pioneering work is usually adopted. Here, the assumptions of this mathematical solution are revisited in order to accurately fit permeation transients. The original formula for $t_T$, which is defined as the interception of the linear asymptote for the rising transient with the $t$-axis, can be expressed as [19]:

$$t_T = \frac{L^2}{6D_L}\left[1 + \frac{3\alpha}{\beta} + \frac{6\alpha}{\beta^2} - \frac{6\alpha}{\beta^3}(1+\beta)\ln(1+\beta)\right] \quad (22)$$

where $t_L$ is defined as $L^2/6D_L$ and the ratios equal: $\alpha = KN_T/N_L$ and $\beta = KC_L^0/N_L$. It must be noted that this solution has been derived assuming an initially empty specimen: $C_L(t=0) = 0$; $C_T(t=0) = 0$, and constant concentration as boundary conditions: $C_L(x=0) = C_L^0$; $C_L(x=L) = 0$. Therefore, errors will inevitably arise when applying this equation to pre-charged specimens – or for consecutive steps as in the present work – and for a constant flux modelling of hydrogen entry.

Since $t_L$ represents the lag time ($t_{0.63}$) without trapping effects, the magnitude $t_T$ can be defined as the apparent lag time, i.e. $t_T = L^2/(6D_{app})$, and represents the experimental $t_{0.63}$ time including trapping effects. Therefore, the relationship between diffusion times can be expressed as:

$$\frac{t_T}{t_L} - 1 = 3K\bar{N}\left[\frac{1}{K\theta_L^0} + \frac{2}{(K\theta_L^0)^2} - \frac{2}{(K\theta_L^0)^3}(1+K\theta_L^0)\ln(1+K\theta_L^0)\right] \quad (23)$$

This function is plotted in Figure 4 for a fixed trap density and different binding energies, i.e. for different $K$ values; Figure 4.a. and 4.b. demonstrate that two asymptotic regimes can be defined, and the curves are just shifted for a different trap density.

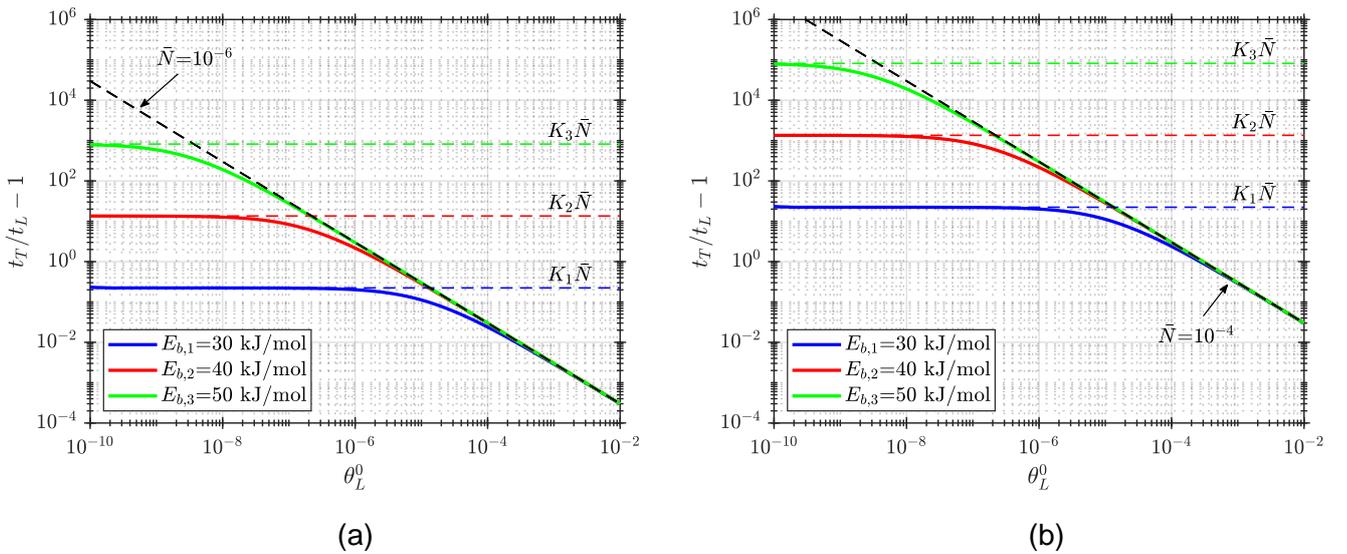

Figure 4. Evolution of $t_T/t_L - 1$ following (23) for different binding energies and asymptotic solutions, for $N_T/N_L$ equal to: (a) 10$^{-6}$ and (b) 10$^{-4}$.

In the saturated region (right part), only the number of traps can be fitted, $\bar{N} = N_T/N_L$; on the contrary, on the diluted region (left part) it is only possible to determine the product $K\bar{N} = KN_T/N_L$. This limiting case was also discussed by Kumnick and Johnson [12]. For the univocal determination of $N_T$ and $K$, a range covering both regimes must be tested; the drawback is that the behaviour turning point is not known a priori. A similar mapping scheme was proposed by Raina et al. [21] in order to determine trapping parameters. For the saturated region $K\theta_L^0 \gg 1$:

$$\frac{t_T}{t_L} - 1 = \frac{3\bar{N}}{\theta_L^0} = \frac{3N_T}{C_L^0} \tag{24}$$

It must be noted that, even though in some works the limitation is not noted, this expression is only applicable, as shown by McNabb and Foster [19] and by Kumnick and Johnson [12], for the limiting case of saturated traps. Additionally, the numerical solution is derived from a two-level equation and for the lattice concentration $C_L^0$ so the inclusion of apparent concentration $C_{app}$ is not numerically consistent. For diluted concentration conditions, i.e. for $K\theta_L^0 \ll 1$ and low trapping occupancy, $\theta_T \ll 1$:

$$\frac{t_T}{t_L} - 1 = K\bar{N} = \frac{KN_T}{N_L} \tag{25}$$

The limiting cases are vital for the correct design of an experimental test program. When apparent diffusivities are fitted using analytic transients instead of the $t_{0.63}$ method, the following equivalence can be considered:

$$\frac{t_T}{t_L} = \frac{D_L}{D_{app}} \tag{26}$$

Trapping characterisation is thus based on the measurement of the deviation from ideal behaviour and requires the previous estimation of lattice diffusivity $D_L$. Two options are possible: (i) to consider that lattice diffusivity represents hydrogen random walk through a bcc iron ideal crystal. In this case, ab initio simulations are useful; and (ii) to assume that lattice diffusivity is represented by the permeation of hydrogen when traps are completely filled; the apparent diffusivity after many permeation transients is then taken as $D_L$. However, for weak traps, this can be hard to achieve even after many transients. Other methodologies based on decaying transients and slope analysis have also been proposed [31]. In the present work, the first alternative is considered for the sake of simplicity and a theoretical value of $D_L = 4.598 \times 10^{-9}$ m²/s at room temperature is taken from the ab initio calculations performed by Jiang and Carter [32].

### 3.4. Hydrogen entry modelling

Two boundary conditions are analysed: (i) constant concentration (CC) in which a constant scale occupancy is imposed on the entry side, $\bar{\theta}_L(x=0)$; (ii) constant flux (CF): in this case, a constant $\bar{j}(x=0)$. The following condition must be fulfilled [33] for the CF model:

$$\bar{j}_{in} = \frac{i_{ss}}{F} = \frac{i_c}{F} - \frac{i_r}{F} \tag{27}$$

where $F$ is the Faraday's constant. The current density represents the experimentally imposed $i_c$ (0.52, 1.04 and 2.08 mA/cm²) and the recombination current $i_r$ is found by considering the experimental steady state output current $i_{ss} = i_p(t \to \infty)$. The output flux in units [atoms/m²/s] is determined by the scale flux $\bar{j}$ calculated in the 1D FE model on the exit side ($x = L$);

$$j_{out} = \frac{D_L C_L^0 N_L}{L} \bar{j}(x = L) = j_{ss} N_L \bar{j}(x = L) \tag{28}$$

In order to compare the FE results with the experimental transients, the numerical output current density [A/m²] is found considering Faraday ($F$) and Avogadro ($N_A$) constants:

$$i_p = \frac{j_{out} F}{N_A} \tag{29}$$

### 3.5. Polycrystal 2D model

Following the previous work of the authors [34], a 2D polycrystal model is used to explicitly simulate grain boundary trapping. Slabs of $L \times 1000$ µm² are modelled, where $L$ is equal to 580 µm for the 925°C/40min sample and 480 µm for 1100°C/5min. The synthetic microstructures are generated using a Voronoi tessellation implemented in a python script plug-in facilitated from [35]; this script has been modified in order to automatically assign different material properties to grains and grain boundaries; in future research, this automatization will be exploited to study diffusion anisotropy and texture effects. The algorithm also calculates every grain area, $A_i$, and the equivalent diameter. $d_i = 2\sqrt{A_i/\pi}$; then, a histogram of $d_i$ distributions is obtained and fitted using a normal distribution, as shown in Figures 5 and 6. A loop of microstructure generation is repeated until the mean $d_i$ values and its standard deviation approximately correspond to the experimentally observed grain size and deviation.

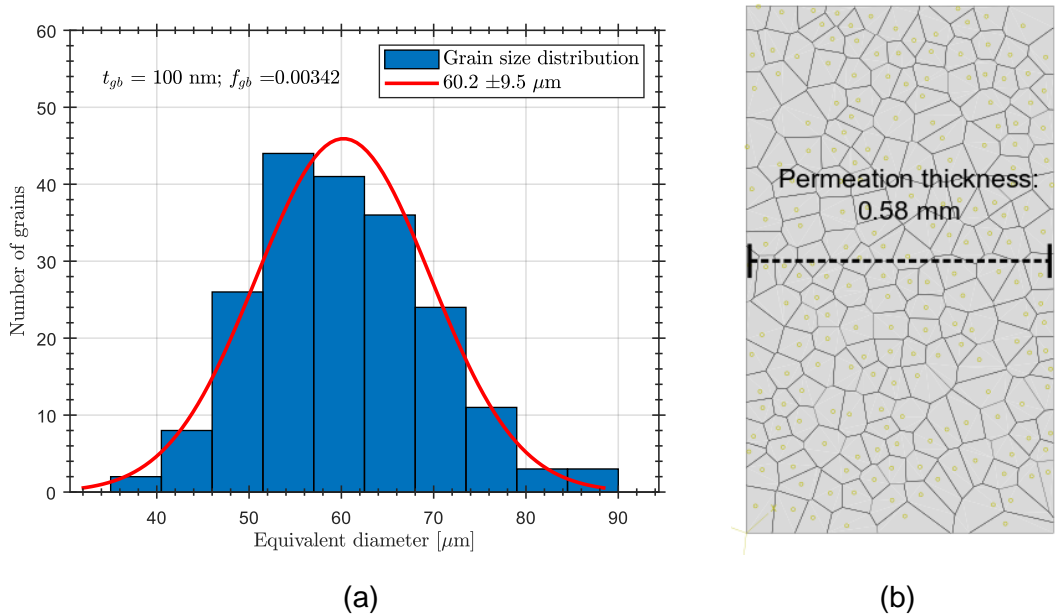

(a)          (b)

Figure 5. Synthetic microstructure generated by Voronoi tessellations to reproduce the sample $T_g = 925°C$ (b), and the corresponding histogram of grain size distribution (a).

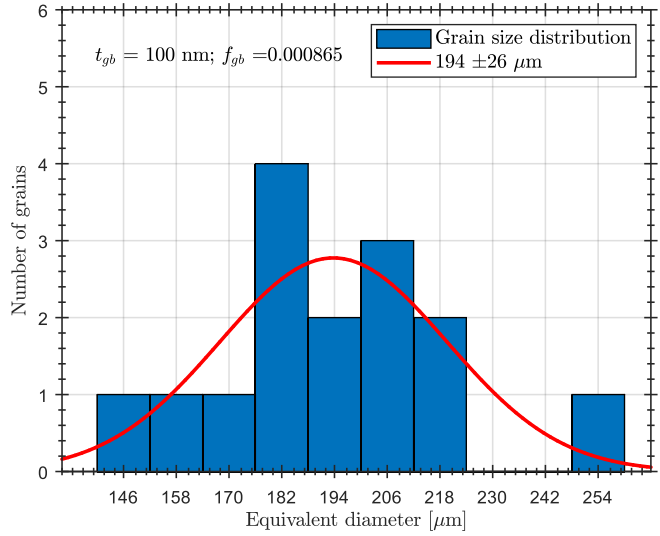 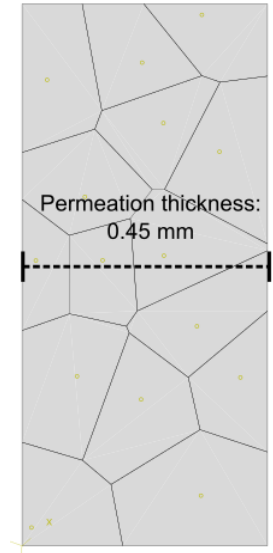

(a) (b)

Figure 6. Synthetic microstructure generated by Voronoi tessellations to reproduce the sample $T_g = 1100°C$ (b), and the corresponding histogram of grain size distribution (a).

A finite grain boundary thickness is considered for modelling trapping; an alternative approach would be based on a zero-thickness interface within a Crystal Plasticity FEM framework in which the evolution of dislocations is calculated for each integration point and related to a local trapping density [36,37]. However, the goal of the present work is to provide a usable model considering the Mass Diffusion module and automatized by python scripts without the need of subroutines so grain boundaries are designed with a given thickness that reproduces a layer where hydrogen is trapped due to the misorientation, geometrically necessary dislocations (GND) [38] and the likely carbon segregation. In the present calculations, an intermediate thickness of $t_{gb} = 100$ nm is considered [34].

Different homogenization techniques can be followed to define an effective diffusivity [39]; here, the upper Hashin-Shtrikman bound is considered. This expression was originally proposed for the determination of elastic moduli of multiphase materials [40] and derived later for multiphase diffusivity [41]:

$$D_{H-S} = D_{gb} + \frac{1 - f_{gb}}{\frac{1}{D_L - D_{gb}} + \frac{f_{gb}}{2D_{gb}}} \qquad (30)$$

where $f_{gb} = A_{gb}/A_{grains}$ is also determined in the generation script; this grain boundary fraction is proportional to the grain boundary thickness but also depends on the boundary network, i.e. the larger grain size, the lower $f_{gb}$. Here it is assumed that this composite diffusivity is equal to the experimentally found apparent diffusivity, i.e. $D_{H-S} = D_{app}$, so a grain boundary diffusivity $D_{gb}$ is iteratively obtained for each model with the corresponding grain boundary fraction. Following this procedure, since $D_{app}$ varies depends on concentration, $D_{gb}$ is implemented in ABAQUS by considering a concentration-dependent diffusivity table. Solubility properties are assigned following the approach from [34]: solubility of grains is taken as 1 and a segregation is defined as the grain boundary solubility introduced in the corresponding material properties. To determine the segregation magnitude, a low lattice occupancy is assumed, $\theta_L \ll 1$, and Equation (9) is rearranged considering $C_{gb} = \theta_T N_T$.

$$C_T = \frac{N_T K}{N_L} C_L \left(\frac{1}{1 + K\theta_L}\right) \qquad (31)$$

Mass diffusion analysis in ABAQUS does not consider concentration-dependent solubilities so the diluted case defined above, $K\theta_L^0 \ll 1$, is assumed and a constant non-dimensional segregation factor can be implemented:

$$s_{gb} = \frac{N_T K}{N_L} \qquad (32)$$

This is an advantage of the polycrystalline model with $D_{gb}$ and $s_{gb}$: the density of traps $N_T$ and the binding energy $E_b$ do not need to be explicitly determined in the diluted case.

Hydrogen input is modelled considering also, as in the 1D FEM and the 1D analytical modelling approaches, two boundary conditions: constant concentration and constant flux. CC model requires a concentration boundary condition whereas the CF in ABAQUS involves a distributed surface flux load. These charging conditions and the fixed exit zero-concentration are only applied to grain surfaces; output flux from grain boundaries is not well understood yet [42,43]. The influence of palladium layer on the exit side is not simulated here.

## 4. Results and discussion

### 4.1. Experimental permeation transients

Following the experimental procedure described in Section 2, galvanostatic conditions are applied to the entry surface, i.e. a fixed current density $i_c$, as shown in Table 1. Three rising steps and three consecutive decaying steps are imposed, obtaining a permeation current on the exit side $i_p$. Permeation times are not the same for every step since the charging current is increased or decrease when a stable $i_{ss}$ is observed. It can be seen that the output current is much lower (µA/cm²) that the input $i_c$, demonstrating that recombination phenomena are significant even though poisoning As$_2$O$_3$ has been introduced in the electrolyte.

| $T_g = 925°C$ | | $i_c$ (mA/cm²) | $i_p$ ($i_0$ / $i_{ss}$) (µA/cm²) | $\Delta i_p$ (µA/cm²) |
|---|---|---|---|---|
| Rise | 1st | 0.52 | 0.0 / 26.0 | 26.0 |
| | 2nd | 1.04 | 26.0 / 50.4 | 24.4 |
| | 3rd | 2.08 | 50.4 / 83.7 | 33.3 |
| Decay | 1st | 1.04 | 83.7 / 47.2 | -36.5 |
| | 2nd | 0.52 | 47.2 / 23.6 | -23.6 |
| | 3rd | 0.0 | 23.6 / 0.0 | -23.6 |
| $T_g = 1100°C$ | | $i_c$ (mA/cm²) | $i_p$ ($i_0$ / $i_{ss}$) (µA/cm²) | $\Delta i_p$ (µA/cm²) |
| Rise | 1st | 0.52 | 0.0 / 59.0 | 59.0 |
| | 2nd | 1.04 | 59.0 / 108.8 | 49.8 |
| | 3rd | 2.08 | 108.8 / 198.4 | 89.6 |
| | 1st | 1.04 | 198.4 / 118.2 | -80.2 |
| Decay | 2nd | 0.52 | 118.2 / 75.4 | -42.8 |
| | 3rd | 0.0 | 75.4 / 0.0 | -75.4 |

Table 1. Input and output currents.

The complete transient is shown in Figure 7 for the sample with a heat treatment of 925ºC for 40 minutes. Figures 8.a. and 8.b. display the separated rising and decaying permeation transients whereas Figures 8.c. and 8.d. show the normalised output current, equivalent to the normalised output flux, that is fitted considering the analytical expressions described in Section 3.2. Visually, it can be concluded that trapping effects are more pronounced during the first rise step because the output current is delayed. This result was expected since at the beginning of permeation traps were completely empty but in the third step traps are occupied to a certain level, so retention effects are weaker. The same physical process is happening in decaying steps: during the first decay trapping sites should be occupied so permeation is faster and at the last step reversible traps have been emptied again.

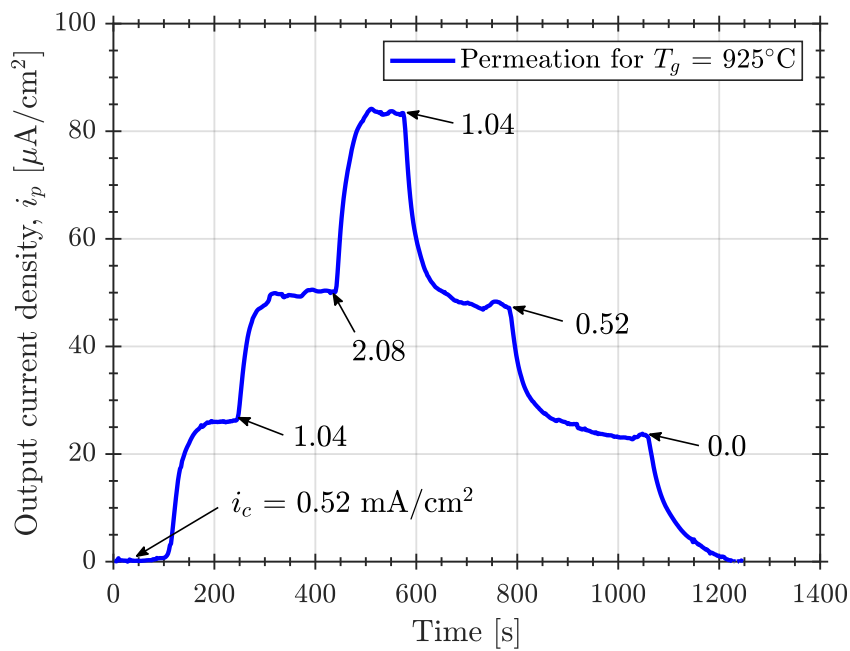

Figure 7. Complete permeation transient with three rising and three decaying steps for the sample heat-treated at 925ºC for 40 minutes.

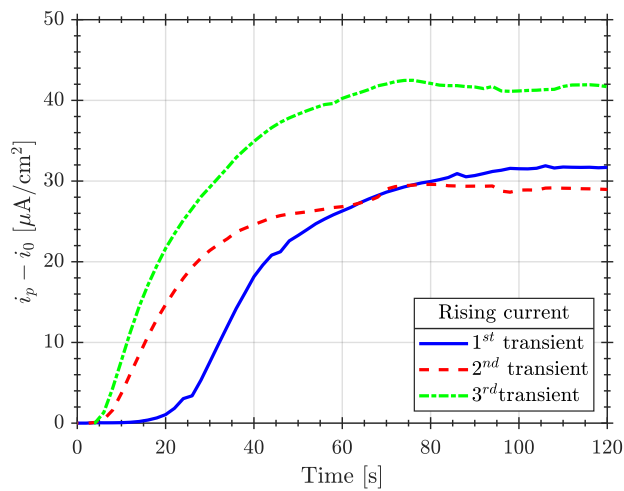
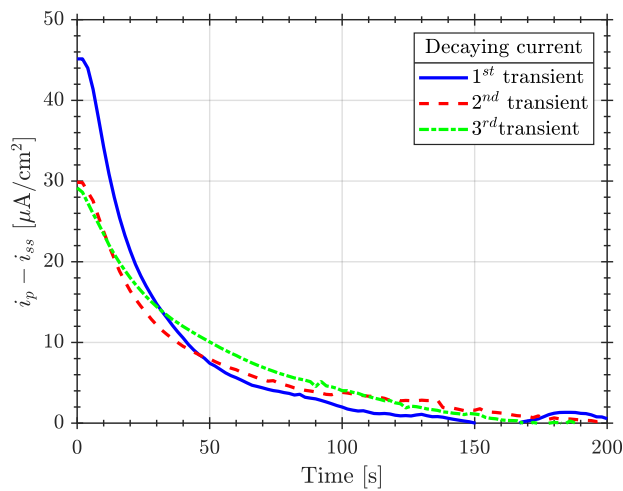

(a)

(b)

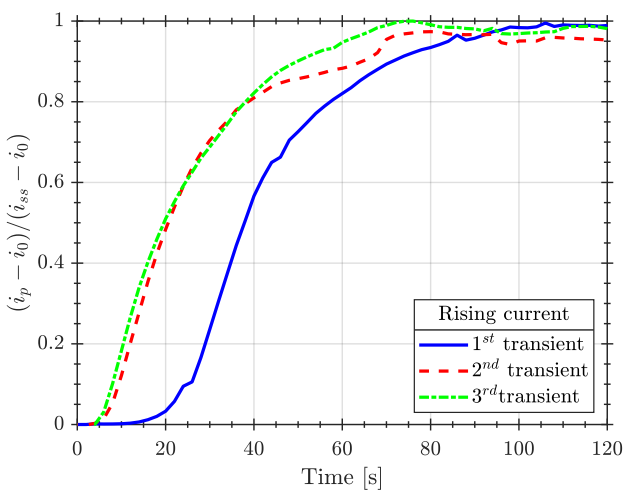
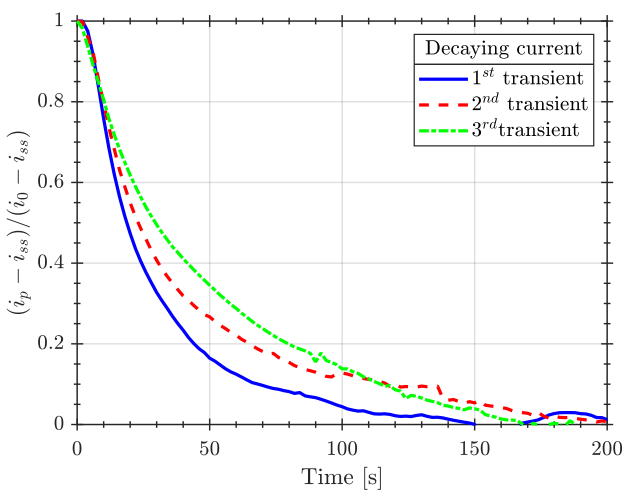

(c)

(d)

Figure 8. Separated permeation transients for $T_g = 925$ºC: (a) dimensional rise steps; (b) dimensional decay steps; (c) normalised rise steps; (d) normalised decay steps.

Similarly, all experimental steps are shown in Figure 9 for the sample with a heat treatment of 1100ºC for 5 minutes. It must be taken into account the different y-axis and x-axis scales. Individual steps are also plotted in the corresponding units in Figures 10.a. and 10.b. and normalised in Figures 10.c. and 10.d. The delaying produced by trapping sites in the first rise and in the last decay are also observed, as discussed above for the 925ºC sample. However, for the 1100ºC sample the difference of trapping effects for the extreme steps in comparison with the intermediate steps (second rise, third rise, second decay and third decay) is more pronounced.

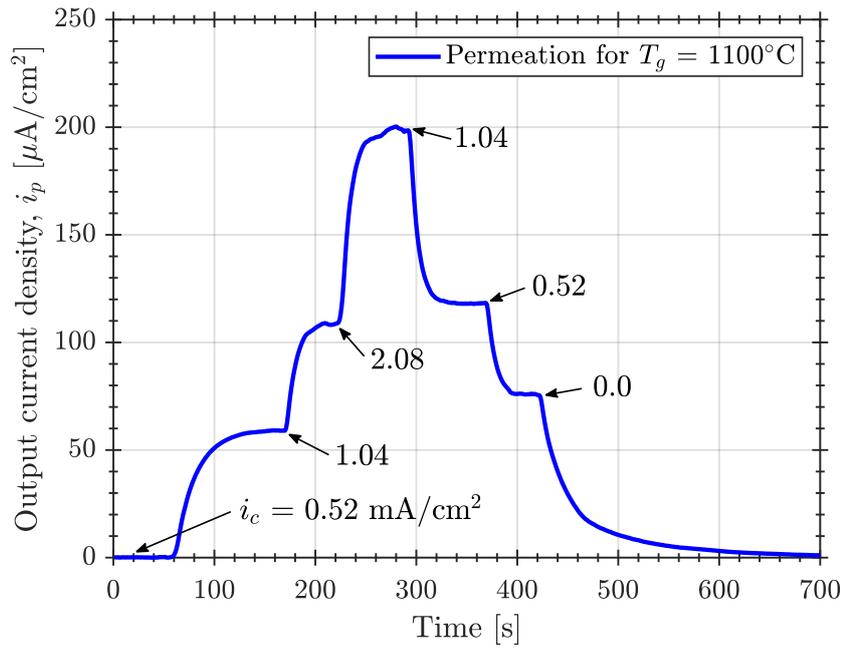

Figure 9. Complete permeation transient with three rising and three decaying steps for the sample heat-treated at 1100ºC for 5 minutes.

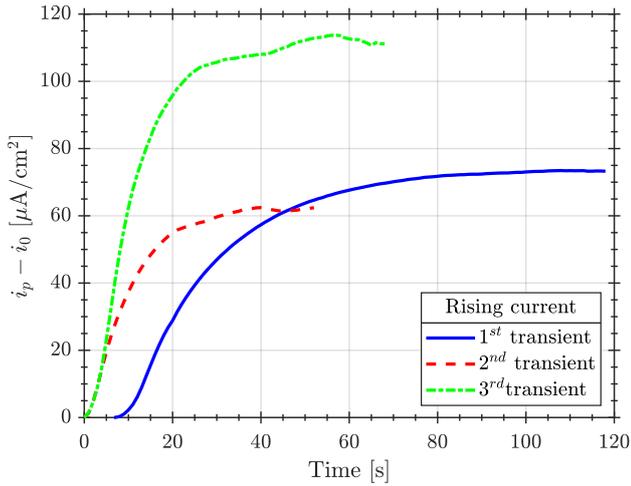
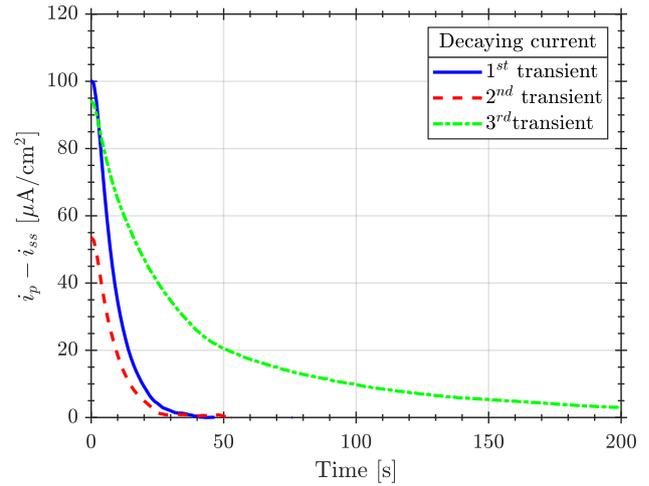

(a)                                    (b)

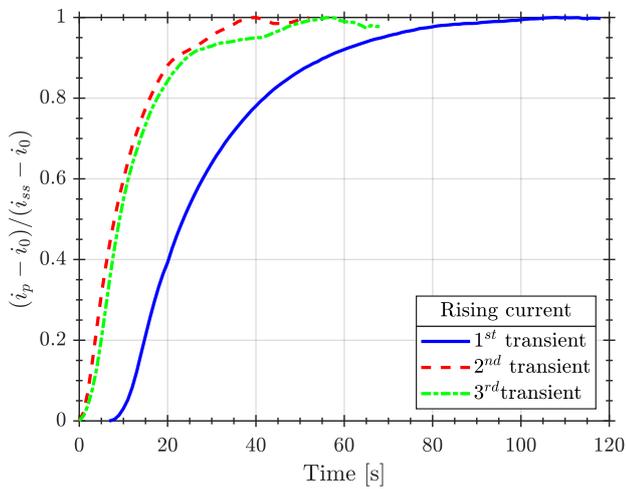
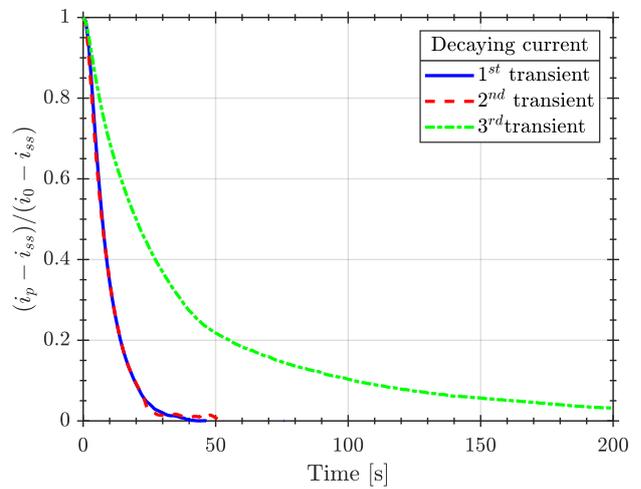

(c)                                    (d)

Figure 10. Separated permeation transients for $T_g = 1100°C$: (a) dimensional rise steps; (b) dimensional decay steps; (c) normalised rise steps; (d) normalised decay steps.

Even though the repeatability of results has not been studied, it is assumed that, for the same sample and the same charging conditions, the scatter in permeation transients would be very small [44]; therefore, tests for two samples after different heat treatments are compared and both permeation transients are shown in Figure 11.a. Permeation for the coarse-grained sample ($T_g = 1100°C$) is faster and the output current value is higher. A weaker trapping effect is explained by the lower expected fraction of grain boundaries due to the coarse grain size; on the other hand, the higher output flux can be due to the lower segregation of hydrogen in grain boundaries. However, experimental results cannot be directly compared because heat treatments produce different thicknesses. To avoid a possible misinterpretation, complete transients are plotted in Figure 11.b. considering a normalised y-axis to the maximum output current of each sample and a normalised time, $\bar{t} = tD_L/L^2$. The time normalisation indicates that the difference of delayed diffusion is partly explained by the different thickness but after normalisation a more pronounced trapping effect is still found for $T_g = 1100°C$.

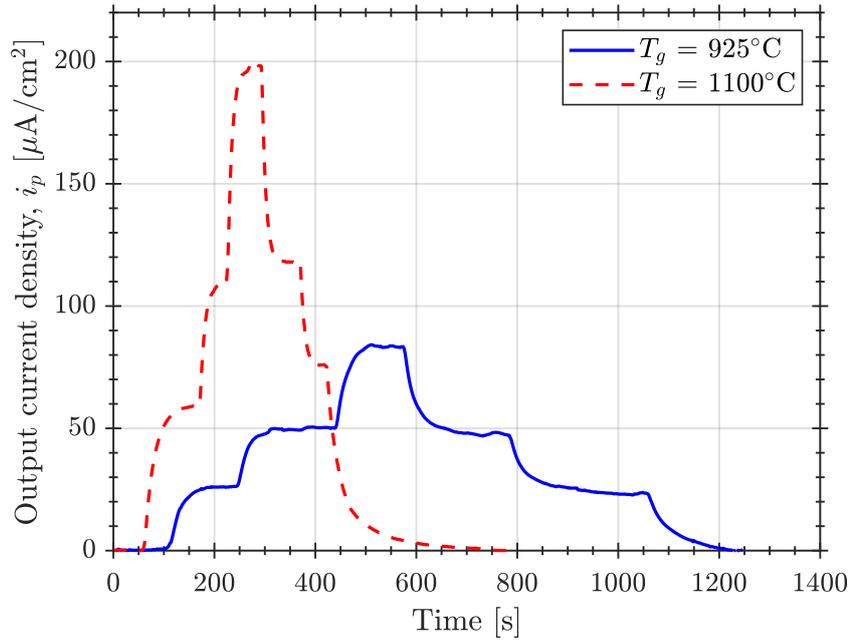

(a)

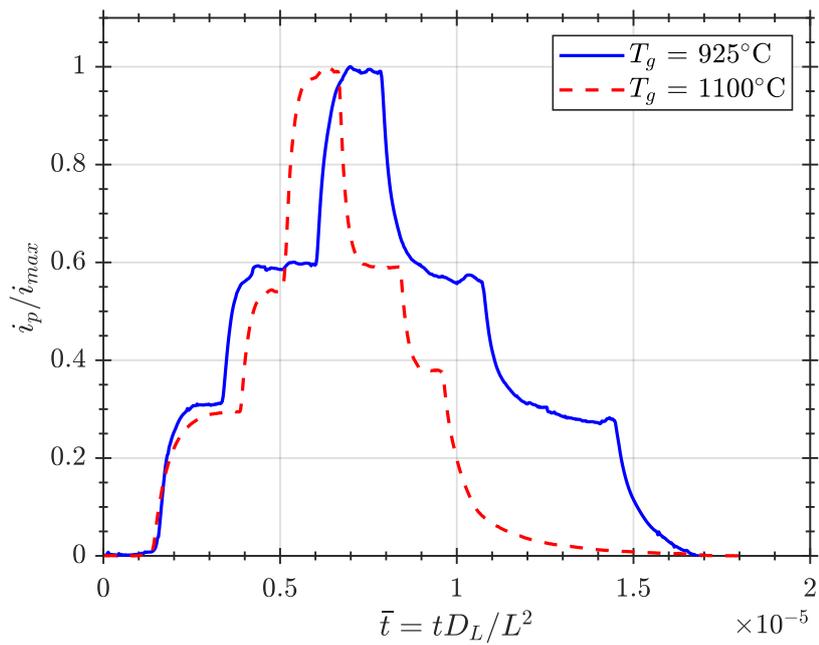

(b)

Figure 11. Comparison of complete permeation transients for $T_g$ = 925⁰C / 40 min and $T_g$ = 1100⁰C / 5 min. (a) dimensional output current and time; (b) normalised current and time.

### *4.2. Apparent diffusivity and concentration*

From the individual transients that have been normalised, the apparent diffusivity is fitted using a non-linear lest squares algorithm implemented in Matlab for both CC and CF analytical solutions considering equations (17) and (20), respectively. A high number of terms for the expansion series are here considered, $n = 50$, due to the good convergence and low computational time required. All fitted curves are shown in Appendix A for every step of each sample.

| $T_g = 925°C$ | | $D_{app}$ (CC fitting: eq. (17)) (µm²/s) | $D_{app}$ ($t_{0.63}$ fitting: eq. (18)(20)) (µm²/s) | $D_{app}$ (CF fitting: eq. (20)) (µm²/s) |
|---|---|---|---|---|
| Rise | 1st | 1243.3 | 1304.6 | 909.59 |
| | 2nd | 2131.2 | 2178.7 | 1510.6 |
| | 3rd | 2265.2 | 2135.7 | 1636.2 |
| Decay | 1st | 2152.2 | 2114.4 | 1510.3 |
| | 2nd | 1636.8 | 1664.7 | 1123.0 |
| | 3rd | 1302.1 | 1217.0 | 925.89 |
| $T_g = 1100°C$ | | $D_{app}$ (CC fitting) (µm²/s) | $D_{app}$ ($t_{0.63}$) (µm²/s) | $D_{app}$ (CF fitting) (µm²/s) |
| Rise | 1st | 1155.4 | 1143.4 | 835.6 |
| | 2nd | 3336.6 | 3143.1 | 2405.9 |
| | 3rd | 2899.1 | 2848.8 | 2072.7 |
| Decay | 1st | 3715.0 | 3566.6 | 2683.2 |
| | 2nd | 3779.0 | 3555.5 | 2732.6 |
| | 3rd | 1154.6 | 1128.2 | 1128.2 |

Table 2. Apparent diffusivity determined by different fitting methods.

It must be noted that the fitting procedure termed as $t_{0.63}$ method is based on the analytical solution assuming constant concentration as a boundary condition; this explains the similar values of $D_{app}$ that have been found for CC and for $t_{0.63}$ methods. Therefore, only the CC and the CF fitting procedures are compared in the following discussion. An apparent concentration is determined by assuming the linear behaviour at steady state:

$$C_{app} = \frac{j_{ss}L}{D_{app}} \qquad (33)$$

| $T_g = 925°C$ | | $C_{app}$ (CC fitting) (mol/m³) | $C_{app}$ (CF fitting) (mol/m³) |
|---|---|---|---|
| Rise | 1st | 1.257 | 1.718 |
| | 2nd | 1.422 | 2.006 |
| | 3rd | 2.221 | 3.075 |
| Decay | 1st | 1.318 | 1.879 |
| | 2nd | 0.867 | 1.263 |
| | 3rd | 0.0 | 0.0 |
| $T_g = 1100°C$ | | $C_{app}$ (CC fitting) (mol/m³) | $C_{app}$ (CF fitting) (mol/m³) |
| Rise | 1st | 2.382 | 3.293 |
| | 2nd | 1.521 | 2.109 |
| | 3rd | 3.192 | 4.464 |
| Decay | 1st | 1.484 | 2.055 |
| | 2nd | 0.931 | 1.287 |
| | 3rd | 0.0 | 0.0 |

Table 3. Comparison of CC and CF predictions of apparent diffusivity obtained from $C_{app} = j_{ss}L/D_{app}$.

### 4.3. Surface phenomena

The sub-surface concentration and the entry flux are not known a priori and can be only predicted by analysing permeated hydrogen at the exit side. As already described in Section 3, two modelling assumptions have been considered: constant concentration and constant flux. These two approaches need to be adapted to the two-level governing PDE that is solved within the Finite Element code; in this equation, the dependent variable is $\theta_L(x,t)$ or, equivalently, $C_L(x,t)$. For the case of constant concentration at the entry surface, the lattice concentration $C_L(x = 0) = C_L^0$ must be found:

$$C_L^0 = \frac{j_{ss}L}{D_L} \tag{34}$$

where $j_{ss}$ is the experimental input; for $D_L$ the two options described above are possible, but here the theoretical value is used [32]. Thus, the constant concentration $C_L^0$ is independent of the fitting method, i.e. does not depend on $D_{app}$.

| $T_g = 925°C$ | | $C_L^0$ (mol/m³) | $C_L^{0*}$ (mol/m³) | $i_c$ (mA/cm²) | $i_r$ ($i_{ss} - i_c$) (mA/cm²) |
|---|---|---|---|---|---|
| Rise | 1st | 0.340 | 0.340 | 0.52 | 0.494 |
|  | 2nd | 0.659 | 0.659 | 1.04 | 0.990 |
|  | 3rd | 1.094 | 1.094 | 2.08 | 1.996 |
| Decay | 1st | 0.617 | 1.094 | 1.04 | 0.993 |
|  | 2nd | 0.309 | 0.617 | 0.52 | 0.496 |
|  | 3rd | 0.0 | 0.309 | 0.0 | 0.0 |
| $T_g = 1100°C$ | | $C_L^0$ (mol/m³) | $C_L^{0*}$ (mol/m³) | $i_c$ (mA/cm²) | $i_r$ ($i_{ss} - i_c$) (mA/cm²) |
| Rise | 1st | 0.598 | 0.598 | 0.52 | 0.491 |
|  | 2nd | 1.104 | 1.104 | 1.04 | 0.931 |
|  | 3rd | 2.012 | 2.012 | 2.08 | 1.881 |
| Decay | 1st | 1.199 | 2.012 | 1.04 | 0.922 |
|  | 2nd | 0.765 | 1.199 | 0.52 | 0.445 |
|  | 3rd | 0.0 | 0.765 | 0.0 | 0.0 |

Table 4. Lattice concentration $C_L^0$ on the entry side; modified $C_L^{0*}$; charging current $i_c$, and recombination current $i_r$ for each step.

However, the significance of $C_L^0$ for decaying steps is hard to interpret; for example, the last decaying step shows a very similar $D_{app}$ than the first rise but $C_L^0 = 0$ for the last transient in which $i_c = 0$. Thus, for the three decaying steps, the considered concentration that influences $D_{app}$ is the $C_L^0$ from the previous step, i.e. $C_{L,i}^{0*} = C_{L,i-1}^0$, as shown in Table 4. This variable better describes the dependency of $D_{app}$ as shown in Figure 12; a plateau is observed and is explained by the fact that traps are almost full and diffusivity approaches to $D_L$.

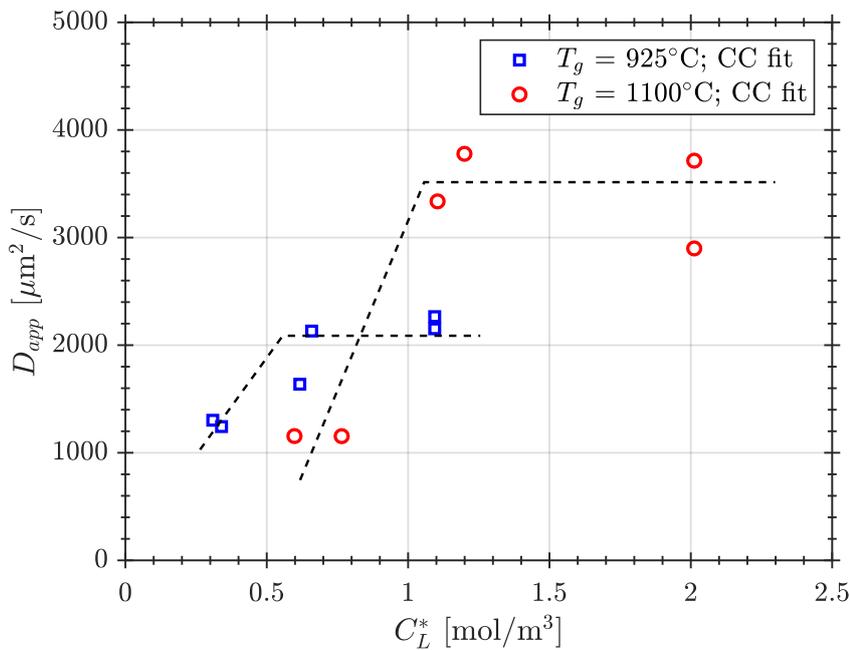

Figure 12. Influence of input lattice concentration $C_L^{0*}$ on the apparent diffusivity.

For the constant flux model, the input flux at each charging step is fixed as the steady state value at the end of that step. Thus, a recombination current can be found by determining the difference between the charging current, $i_c$, and the obtained $i_{ss}$.

$$j_{in} = -D_L \frac{\partial C_L}{\partial x}\bigg|_{x=0} = \frac{i_{ss}}{F} = \frac{i_c}{F} - \frac{i_r}{F} \tag{35}$$

In order to analyse the charging efficiency, the experimental steady state values are plotted against the charging current in Figure 13.a.; it is observed a linear trend and a higher steady state flux for the coarser microstructure, which is explained by the lower trapping effects. In order to discard thickness effects, the steady state flux is multiplied by the thickness in Figure 13.b., following [30]. Some authors [30] have fitted instead a linear relationship between $i_{ss}$ and $\sqrt{i_c}$, which is plotted in Figure 13.c.; this latter dependency should be observed when $i_{ss} \ll i_r$, as is the case [33,45]. However, the most general and valid relationship is the $\sqrt{i_r}$ versus $i_{cc}$, as plotted in Figure 13.d., whose linear dependency is also demonstrated.

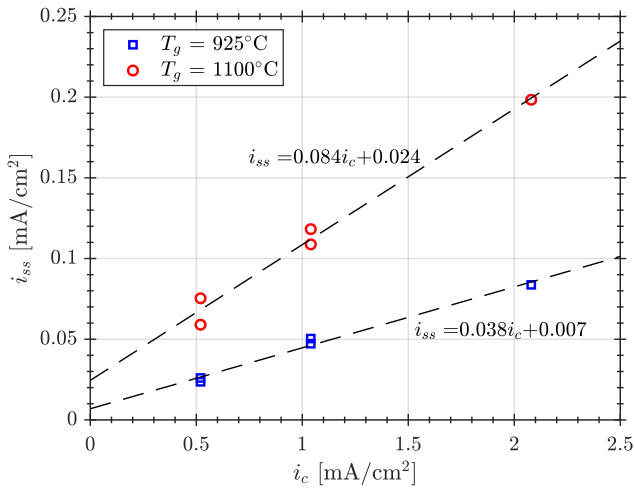

(a)

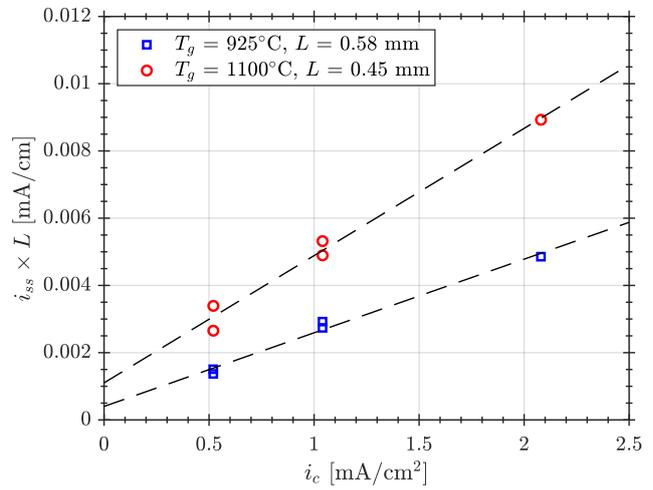

(b)

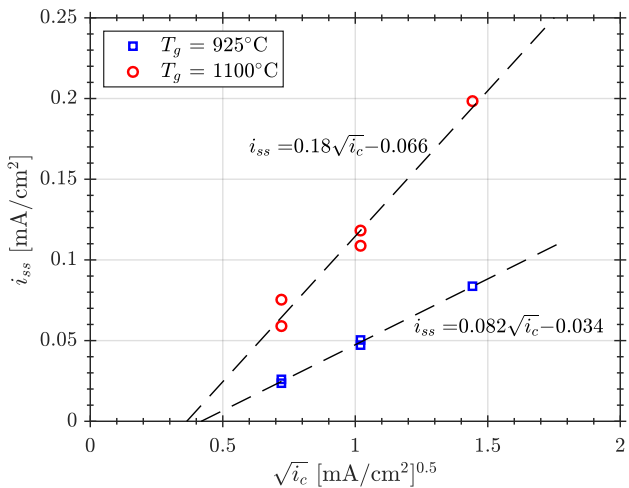

(c)

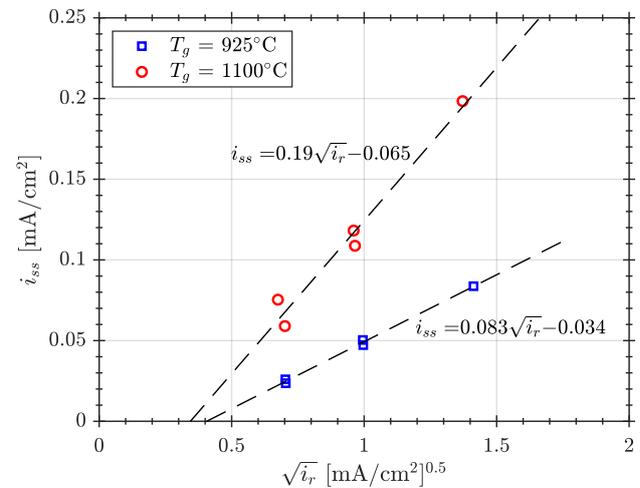

(d)

Figure 13. Dependency of steady state flux on charging and recombination currents.

Whether this higher charging current promotes a higher input concentration is assessed in Figure 14. Both apparent concentration $C_{app}$ and lattice hydrogen concentration $C_L^0$ are plotted for every step and for both samples, $T_g = 925°C$ and 1100°C. As expected from the steady state $i_{ss}$, more hydrogen is being produced in the surface of the coarser microstructure. A linear relationship can be fitted for the evolution of both concentration magnitudes as a function of $i_c$. However, the $C_{app}$ for the first rise lies outside this behaviour; this is explained by the initially empty traps that promote a stronger diffusion delay so the apparent diffusivity is much lower and the corresponding $C_{app} = j_{ss}L/D_{app}$ is much higher.

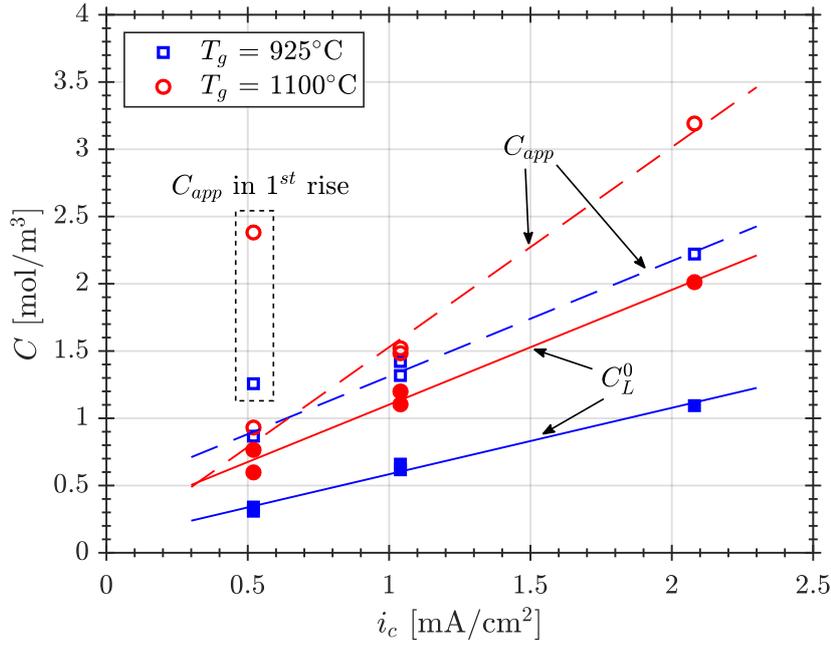

Figure 14. Influence of charging current density on hydrogen apparent and lattice concentrations; for the determination of $C_{app}$ the CC fitting has been considered.

### 4.4. Trapping features

A mapping approach, as described in Section 3.3, is shown in Figure 15. It must be noted that the plotted asymptotes, as already discussed, are only valid for constant concentration. Here, the x-axis corresponds to lattice occupancy and experimental results are included considering the modified concentration, $\theta_L^0 = C_L^{0*}/N_L$, as explained above, i.e. taking the input concentration of previous step for decaying transients.

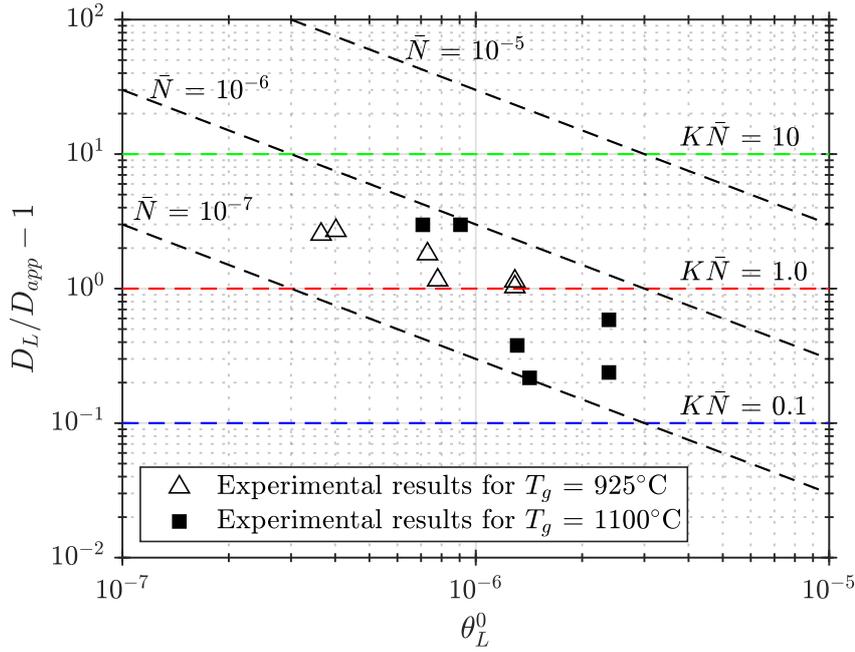

Figure 15. Mapping for CC fitting.

It can be deduced from Figure 15 that $\bar{N}$ lies between $10^{-7}$ and $10^{-6}$; at the same time, $K\bar{N}$ must be at least 2.0. Assuming the theoretical value of $N_L$ = 5.1×10²⁹ sites/m³, the following ranges are determined:

- 5.1×10²² < $N_T$ < 5.1×10²³ traps/m³
- $K$ > 2×10⁶, which is equivalent to $E_b$ > 35.3 kJ/mol at $T$ = 293 K.

Even though a broader $\theta_L^0$ range is necessary to accurately determine $N_T$ and $E_b$, e.g.. by broadening the experimental range of $i_c$, in the following subsections the saturated and diluted assumptions are explored and discussed in order to characterise trapping parameters.

### 4.4.1. Saturated trap assumption

In the saturated regime, the condition $K\theta_L^0 \gg 1$ must be fulfilled. This can be attained when traps are very energetic or when the charging conditions introduce a high amount of lattice hydrogen. In this case, Equation (24) can be used to find $N_T$. It must be recalled that $t_T/t_L = D_L/D_{app}$. It can be seen in Figure 16 that the trap density $N_T$ takes a higher value for the first rise and for the last decay; this fact confirms that the saturated simplification can only be assumed for the intermediate steps. Thus, the $N_T/N_L$ ratio for each sample is defined as the average of these intermediate steps in which traps should be occupied.

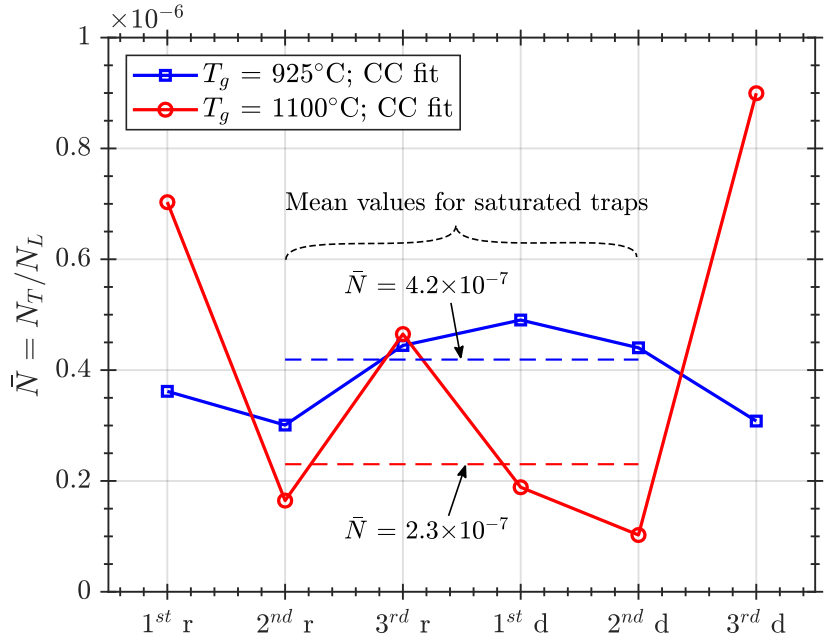

Figure 16. $\bar{N} = N_T/N_L$ for each step and mean values for intermediate transients; in the x-axis, "$r$" represents a rise step and "$d$" a decay step.

### 4.4.2. Diluted trap assumption

In contrast to the saturated regime, the energy of traps can only be fitted when the experimentally ratio $D_L/D_{app}$ depends on $K$, i.e. on the diluted regime. Thus, the product $K\bar{N}$ is determined for each step following Equation (25), but the average is considering only including the extreme steps, i.e. the first rise and the third decay, as shown in Figure 17.

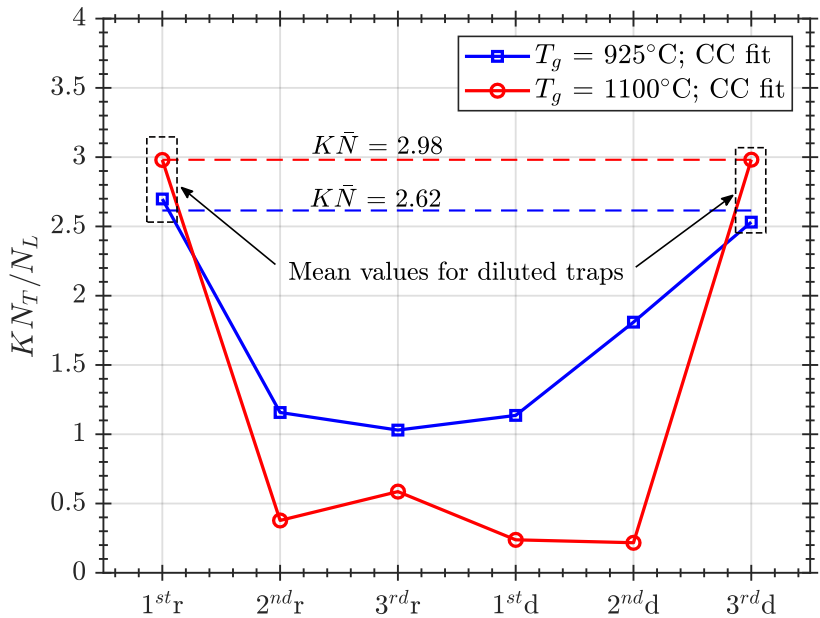

Figure 17. $K\bar{N}$ for each step and mean values for extreme transients; in the x-axis, "$r$" represents a rise step and "$d$" a decay step.

Then, considering the theoretical value $N_L = 5.1 \times 10^{29}$ sites/m³, $N_T$ is determined. Once $N_T$ is found from the saturated assumption in intermediate steps, $K$ can be deduced from $K\bar{N}$ values that have been calculated using the diluted simplification in extreme steps and $E_b$ is obtained assuming $T = 293$ K. These results are shown in Table 5.

|  | $N_T$ (traps/m³) | $E_b$ (kJ/mol) |
|---|---|---|
| $T_g = 925°C$ | $2.13 \times 10^{23}$ | 38.1 |
| $T_g = 1100°C$ | $1.17 \times 10^{23}$ | 39.9 |

Table 5. Characteristic trapping features considering $D_{app}$ fitted from the CC expression.

This fitting methodology is repeated for the $D_{app}$ values obtained using the CF assumption. It must be highlighted that this $D_{app}$ must be taken into consideration since the experimental charging has been performed using a constant current density $i_c$ so galvanostatic conditions have been followed. Nevertheless, as already mentioned, asymptotic expressions to find $N_T$ and $E_b$ have been derived from the permeation numerical solution expressed in Equation (23), which was found by McNabb and Foster [19] assuming a fixed input concentration. This fact limits the application of the present methodology so the parameters shown in Table 6 should be considered with care. However, trap densities are consistent with the usual range of parameters found for pure iron without deformation [12,46]. The comparison of binding energies is more difficult due to the common experimental scatter [12,47,48].

|  | $N_T$ (traps/m³) | $E_b$ (kJ/mol) |
|---|---|---|
| $T_g = 925°C$ | $3.75 \times 10^{23}$ | 37.8 |
| $T_g = 1100°C$ | $2.87 \times 10^{23}$ | 38.3 |

Table 6. Characteristic trapping features considering $D_{app}$ fitted from the CF expression.

### 4.5. Finite Element 1D simulations

The scaled governing PDE (13) is solved in Comsol Multiphysics considering a 1D model with a 1000-node mesh and a geometric bias in order to concentrate elements near the exit node, where a higher accuracy is required for registering the output flux. It has been verified that results are mesh-independent. Permeation time is also divided in 1000 time points and a MUMPS solver is used. Boundary conditions, CC and CF, are implemented considering the Dirichlet and Neumann conditions, respectively, for the associated PDE.

#### 4.5.1. Trapping influence

The values of $N_T$ and $E_b$ that have been experimentally found and shown in Table 5 and 6 are implemented in the 1D FE model. Thus, four situations for FE permeation are simulated:

- CC boundary conditions and $N_T, E_b$ fitted from CC expression.
- CF boundary conditions and $N_T, E_b$ fitted from CC expression.
- CC boundary conditions and $N_T, E_b$ fitted from CF expression.
- CF boundary conditions and $N_T, E_b$ fitted from CF expression.

Since the distribution of lattice concentration $C_L(x)$ is completely linear when steady state fluxes are reached, the numerical $j_{ss}$ coincides completely with the experimental magnitudes. This perfect matching was expected for this 1D model because the imposed $C_L^0 = j_{ss}L/D_L$ in CC approach and the $j_{in} = j_{ss}$ in CF approach are derived from the $j_{ss}$ experimental flux.

On the other hand, the transient slope behaviour in FE results is hard to analyse. Figure 18.a. shows for $T_g = 925°C$ that the CC model better reproduces experimental transients for the rising steps, whereas the CF fits better the decaying steps. Results for $T_g = 1100°C$ (Figure 18.b.) are more consistent because CC predictions are accurate on intermediate steps but CF predictions are better on the first rise and last decay; this is explained by the fact that CF model implies a slower permeation, which occurs when traps are empty and when the surface concentration is lower.

In order to interpret the influence of hydrogen concentration and trapping occupancy, these magnitudes are plotted for the input node, $C_L(x = 0) = C_L^0$ and $\theta_T(x = 0) = \theta_T^0$. The step evolution of $C_L^0$ for the CC model produces a faster hydrogen permeation whereas in the CF model a progressive build-up of hydrogen concentration occurs in the entry side so the permeation is slower. This critical difference explains the different behaviours; a generalised boundary condition, as implemented in different works [4,49], can be more realistic than the limiting cases here studied. When plotting trap occupancy, it is confirmed that traps are nearly fully occupied: $\theta_T^0 > 0.7$ for all steps excluding the last decay transient in which traps are progressively emptied for the CF model but an instantaneous $\theta_T^0$ is imposed for the CC because equilibrium is assumed. This result validates the saturated trap assumption that has been used to determine the trap density $N_T$ but the diluted simplification is limited even for the first rise and for the last decay.

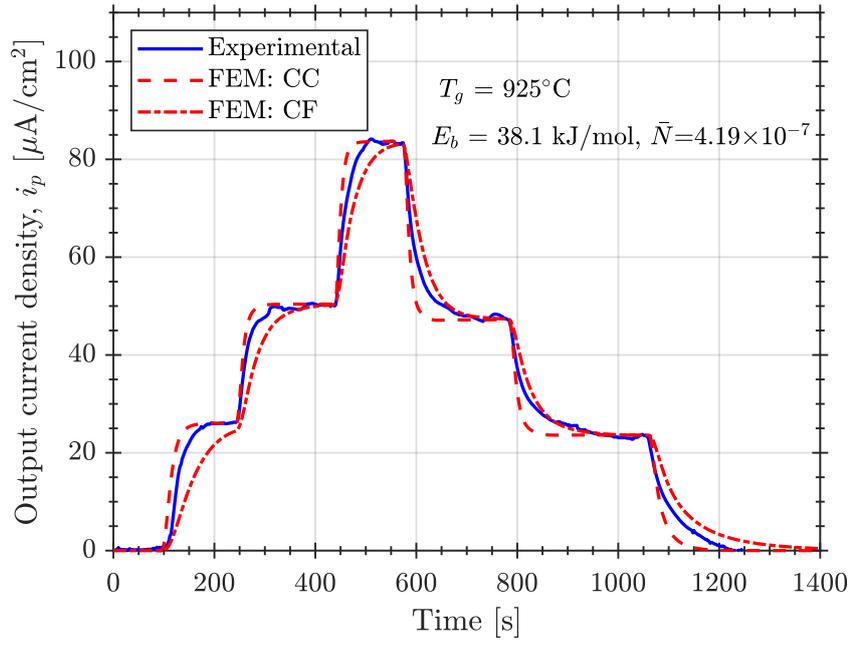

(a)

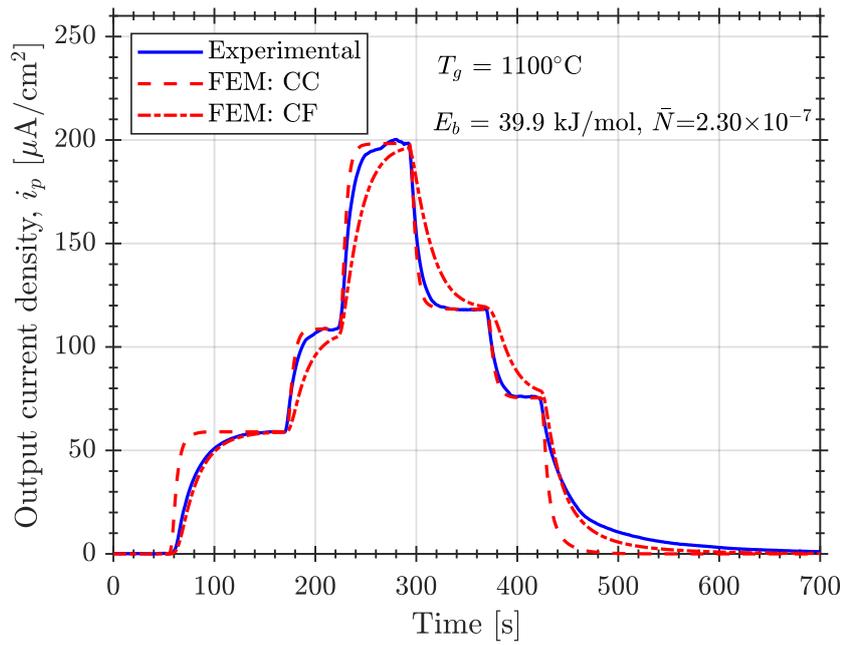

(b)

Figure 18. Comparison of experimental transients and FE results considering trapping parameters determined using $D_{app}$ from CC fitting.

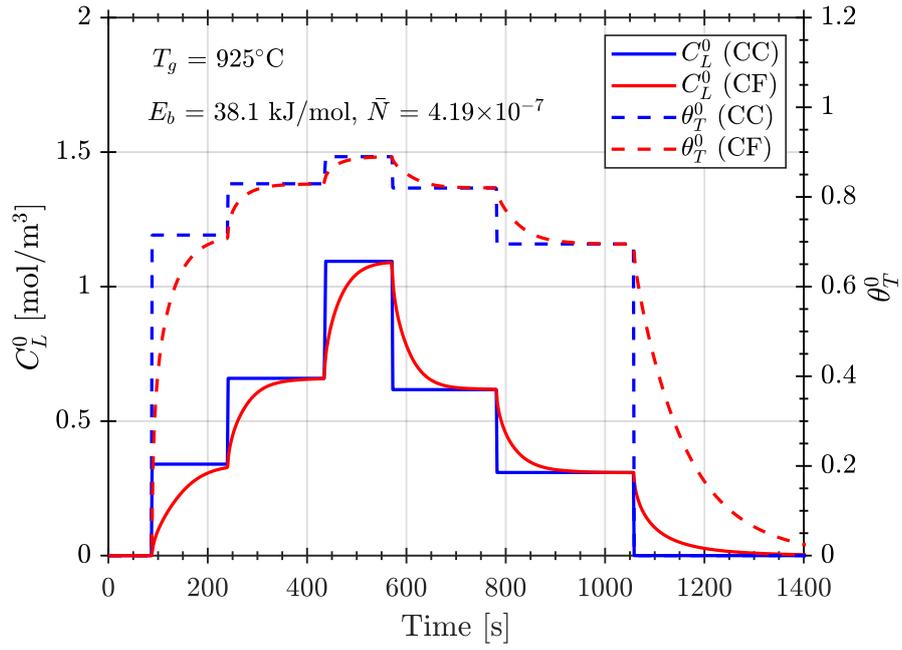

(a)

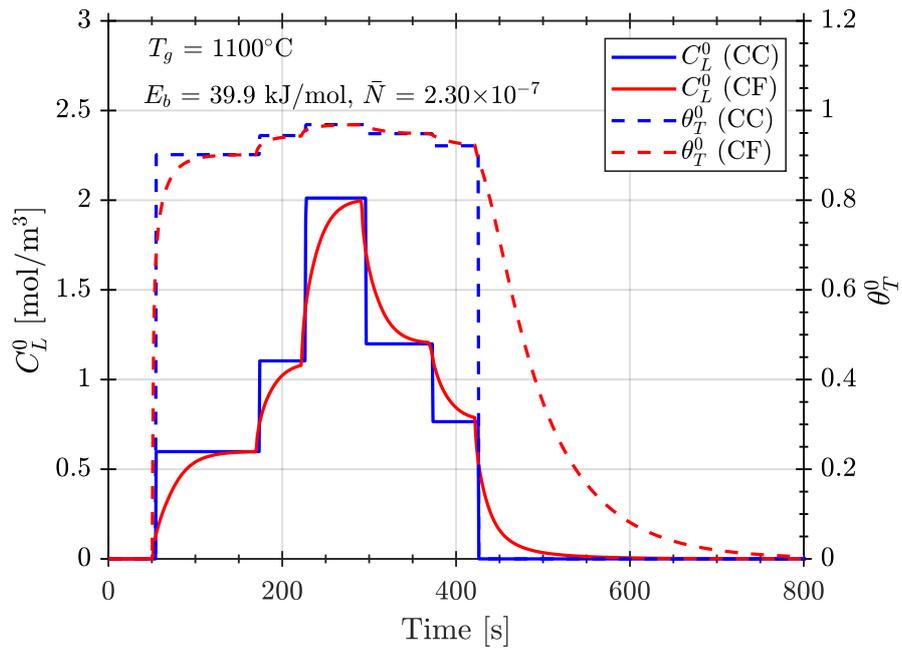

(b)

Figure 19. Numerical evolution of hydrogen lattice concentration $C_L^0$ and trap occupancy $\theta_T^0$ on the entry side considering trrapping parameters determined using $D_{app}$ from CC fitting.

A similar result is obtained for the trapping parameters ($N_T$, $E_b$) that have been determined considering $D_{app}$ fitted using the CF analytical expression. Figure 20.a. shows that the CF model for $T_g$ = 925ºC predicts transients highly deviated from the experimental results, especially for rising steps; Figure 20.b., for $T_g$ = 1100ºC, confirms that CF only matches the experimental results for a diluted concentration, i.e. during the last decay, despite the fact that trapping parameters have been determined using the $D_{app}$ from CF fitting. Thus, it can be concluded that the method of $D_{app}$ calculation is not critical for 1D FE modelling.

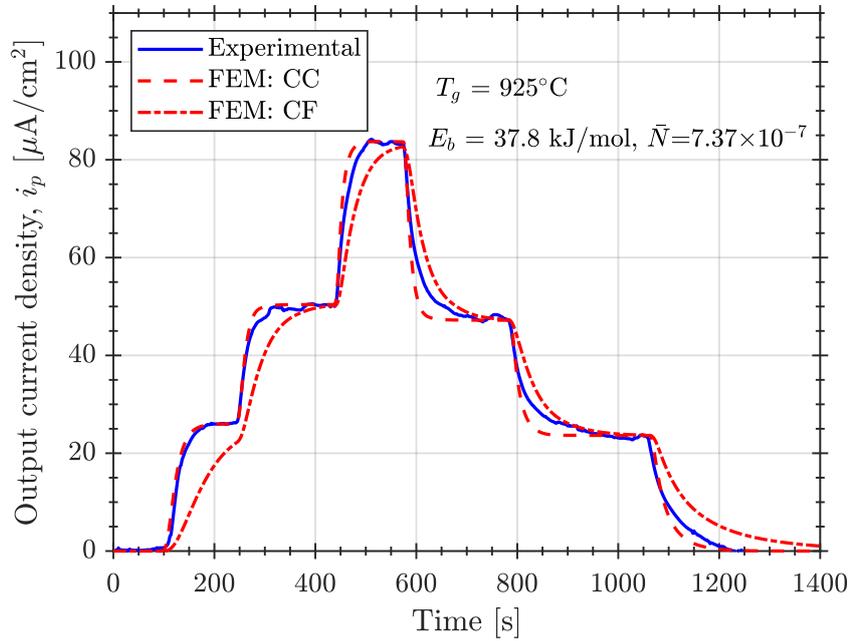

(a)

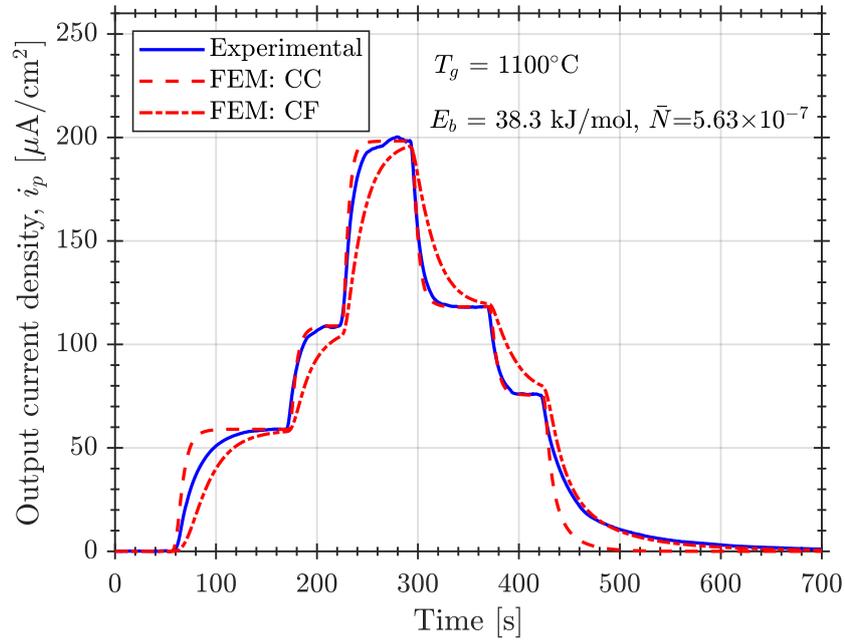

(b)

Figure 20. Comparison of experimental transients and FE results considering trapping parameters determined using $D_{app}$ from CF fitting.

### 4.6. Finite Element polycrystalline model

Diffusivity within grains is assumed as the theoretical $D_L = 4598$ μm²/s [32]. Grain boundary diffusivity, $D_{gb}$, is determined from the $D_{app}$ values that have been found using the CC analytical expression and iterating in Equation (30) for $D_{H-S} = D_{app}$. From the diluted assumption, $s_{gb} = \bar{N}K$, Segregation takes a value of $s_{gb} = 2.62$ for $T_g = 925°C$, and $s_{gb} = 2.98$ for $T_g = 1100°C$, as shown in Figure 17.

Numerical transients obtained from the polycrystal FE model are plotted in Figures 21.a. and 21.b. and compared with the experimental output currents. The output flux for the polycrystal FE model has been integrated over the whole exit surface considering the mass flux in each integration point and the corresponding point area. In contrast to the 1D FE model, the polycrystal predicted transients do not reach the same steady state flux at each step than the experimental $j_{ss}$, even though the same boundary conditions $C_L^0$ and $j_{in}$ have been applied to the entry surface. This can be rationalised because the steady state distributions are not perfectly linear due to the non-homogeneity of the material. Figure 22 plots the distribution of hydrogen concentration in a certain cross-sectional path; the segregation at grain boundaries can be clearly observed. These distributions have been plotted for 925°C at $t = 544$ s and for 1100°C at $t = 289$ s, i.e. near the steady state achievement of the third rise step. The output integrated flux depends on the slope $dC/dx$ that occurs within the grains at the exit surface. For the sample corresponding to $T_g = 1100°C$, the deviation from steady state flux values is lower because the segregation is taking place at a smaller number of boundaries, so the concentration distribution is more similar to the 1D model. For both microstructures, coarse and fine grain size, it is observed that the CF model better represents the experimentally tested permeation.

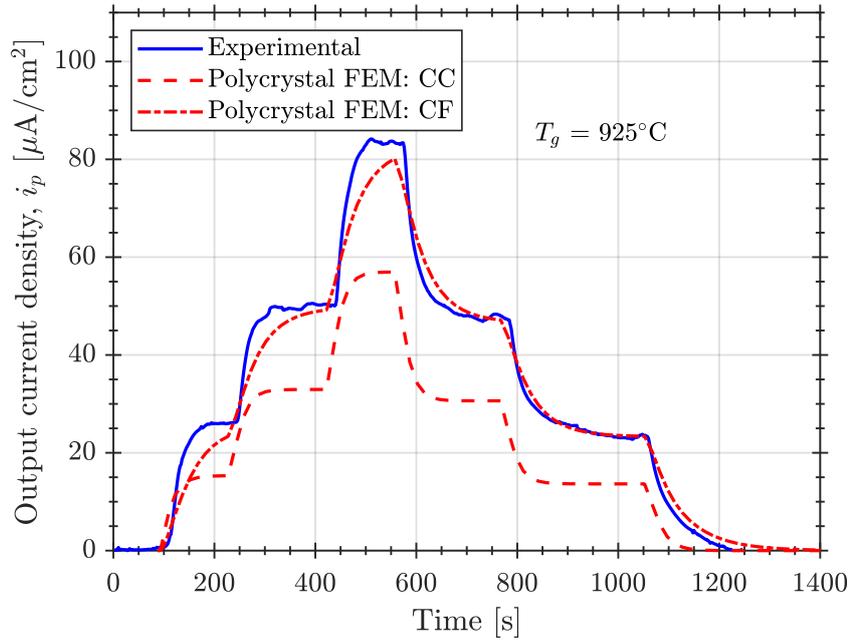

(a)

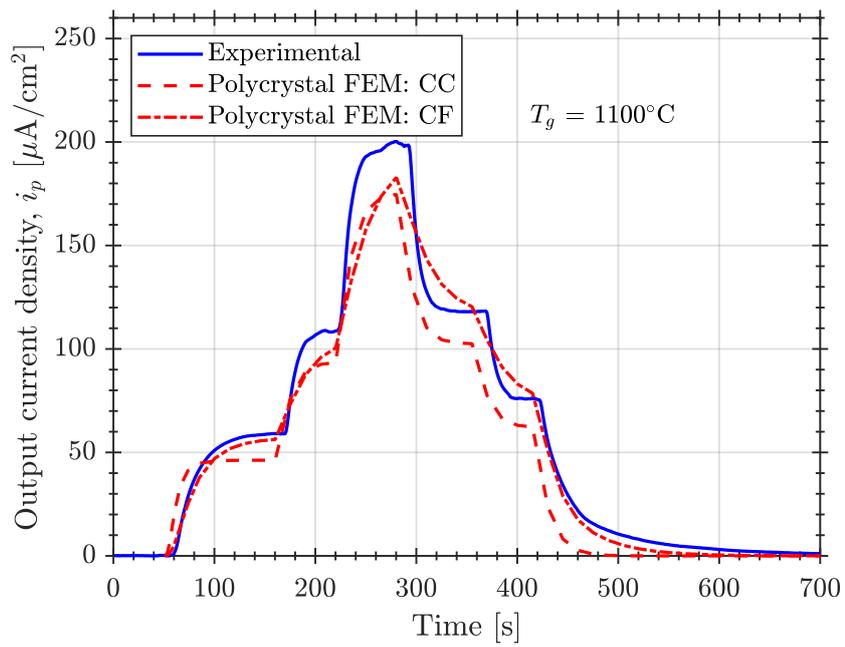

(b)

Figure 21. Comparison of experimental transients and FE results considering $D_{gb}$ and $s_{gb}$ parameters determined using $D_{app}$ from CC fitting.

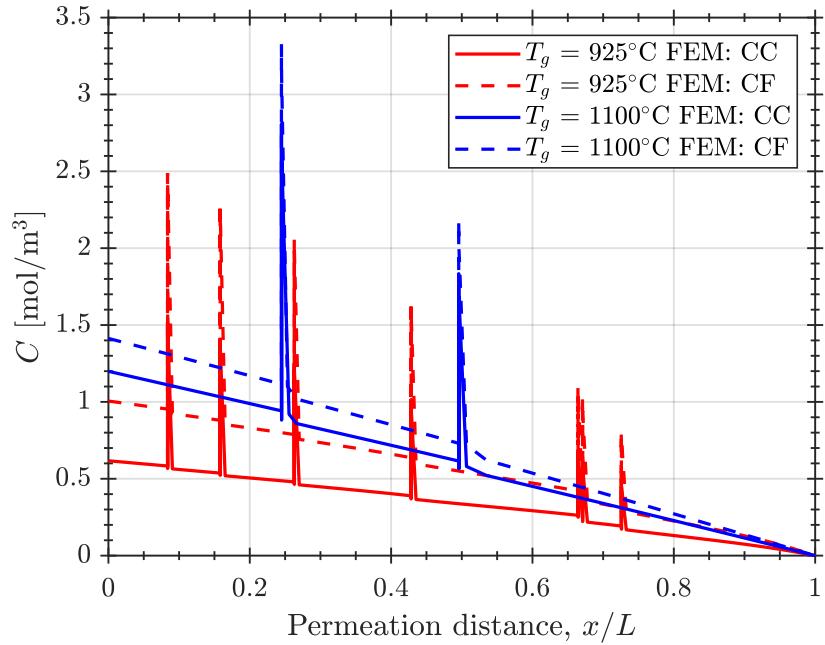

Figure 22. Hydrogen concentration at the maximum rise flux in a cross-section path for CC and CF models. Distribution for 925ºC is plotted at $t = 544$ s and for 1100ºC at $t = 289$ s.

## 5. Conclusions

A numerical methodology has been presented in this work with the aim of analysing trapping effects during hydrogen permeation through metals. In order to assess grain boundary trapping and study grain size influence on trapping phenomena, pure iron has been chosen as the studied material and two heat treatments have been carried out. Considering the relationship between apparent diffusivities that have been analytically fitted and the trapping behaviour, some assumptions are considered to determine $N_T$ and $E_b$. The limiting cases, i.e. saturated and diluted traps, are demonstrated to be helpful in a stepwise permeation test. After this characterisation for both microstructures, two Finite Element approaches have been explored: (i) a 1D permeation model in which trapping effects are reproduced including a trap density and a binding energy; (ii) a polycrystalline model in which trapping is explicitly simulated in grain boundaries by assigning a lower diffusivity and a segregation factor. The following particular conclusions of this methodology can be summarised:

- Trapping sites in pure iron samples have been characterised as medium-energy defects with binding energies between 37.8 and 39.9 kJ/mol; trap densities have been found to take low values, from 1.17×10$^{23}$ to 3.75×10$^{23}$ sites/m$^3$. A stepwise permeation test has been performed in order to evaluate different diffusion regimes.
- A common method to fit binding energies and trap densities, which is based on the asymptotic solution proposed by McNabb and Foster, has been revisited. It has been concluded that only a mapping methodology covering a wide range on concentrations is able to univocally determine at the same time binding energy and trap density values.
- Considering the saturated limiting case, i.e. traps are almost filled, trap densities can be directly obtained. On the other hand, assuming the diluted limiting case,

- i.e. traps are almost empty, both trap density and binding energy influence Finite Element simulations have demonstrated that traps in the pure iron samples are almost filled after the first permeation step but the diluted case can be applied to the first rise and to the last decay transients.
- Due to the fact that McNabb and Foster's solution is based on a constant concentration as a boundary condition on the entry side, the influence of surface phenomena on hydrogen entry has been analysed. Relationships between recombination and charging currents have been fitted, showing that steady state hydrogen flux is much lower than both charging and recombination fluxes; on the other hand, the relationship between apparent concentrations on the entry side and charging conditions only can be indirectly fitted because it depends on the fitting assumptions to determine apparent diffusivities.
- The influence of grain size has been studied by analysing two samples with different heat treatments. Following the analytical fitting of apparent diffusivities, it is concluded that hydrogen diffuses faster through the coarser grain microstructure, so grain boundaries enhance hydrogen trapping, as expected. No acceleration or short circuit effects through grain boundaries are found. The methodology based on McNabb and Foster's asymptotic solution confirms that the coarser grain microstructure implies a higher trap density. However, differences are small and lie on the experimental scatter range. Thus, a polycrystalline model is used to explicitly simulate segregation and diffusion delay in grain boundaries during hydrogen permeation.

Due to the limitations in the determination of trapping features, in future research a broader charging range will be tested in order to complete a mapping and univocally determine trapping parameters. Additionally, generalised boundary conditions will be considered to reproduce realistically the influence of overpotential, charging current and pH. The polycrystalline modelling approach will also be developed to incorporate grain orientations in order to assess possible texture effects and to study the influence of grain boundary nature on trapping phenomena.

ACKNOWLEDGEMENTS

The authors gratefully acknowledge financial support from the Ministry of Economy and Competitiveness of Spain through gran RTI2018-096070-B-C33. E. Martínez-Pañeda also acknowledges financial support from EPSRC funding under grant No. EP/R010161/1 and from the UKCRIC Coordination Node, EPSRC grant number EP/R017727/1, which funds UKCRIC's ongoing coordination.

# Appendix A. Fitting CC vs GF

## A.1. For 925°C

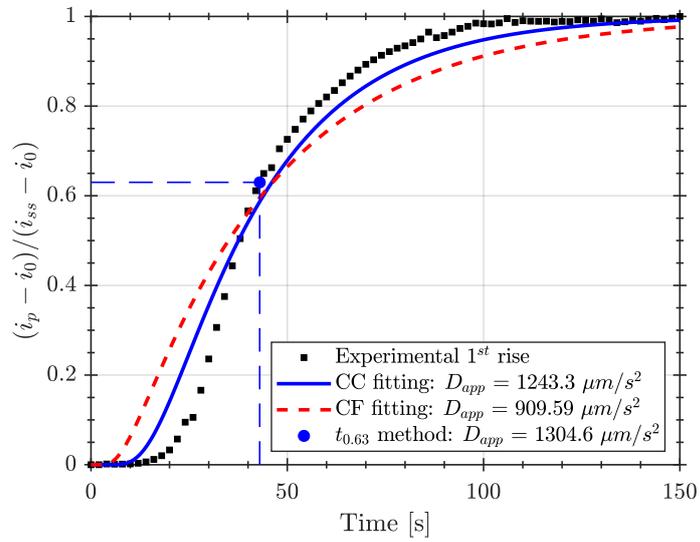

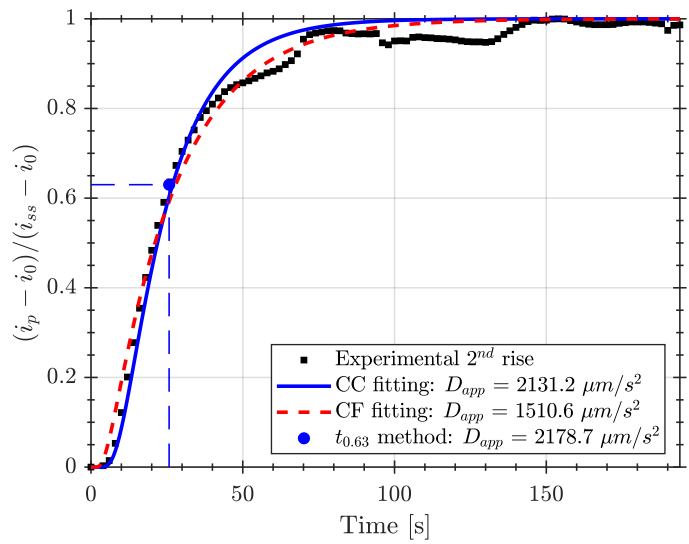

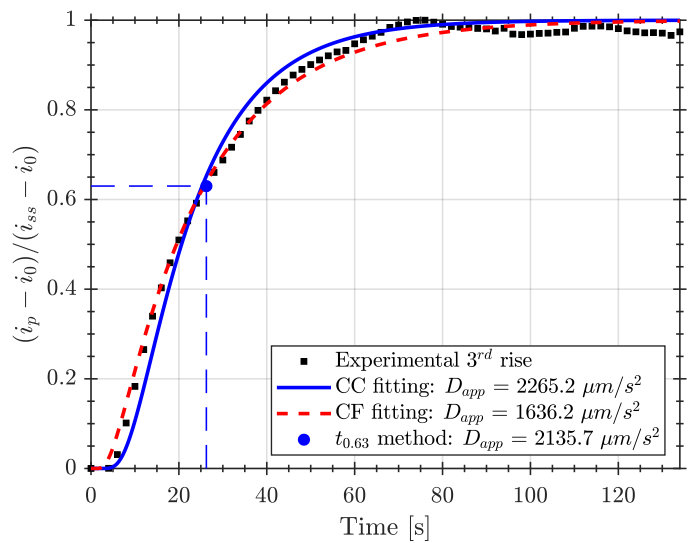

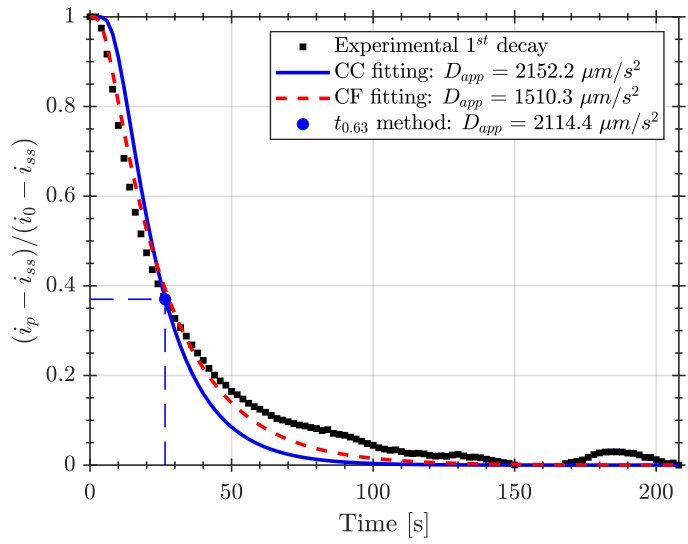

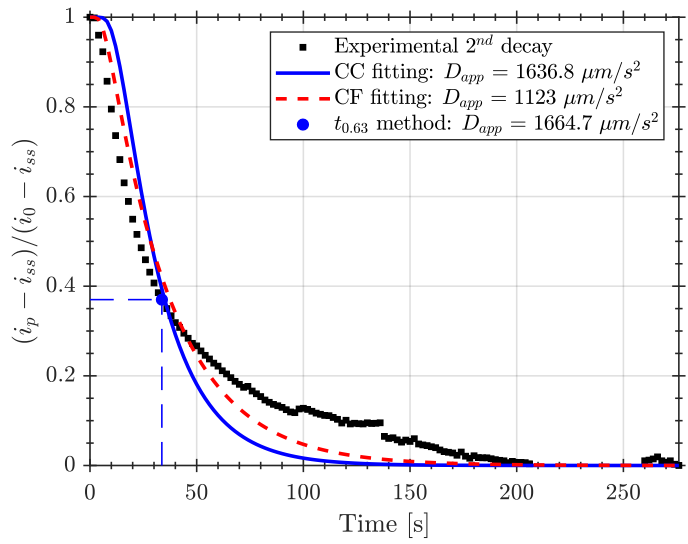

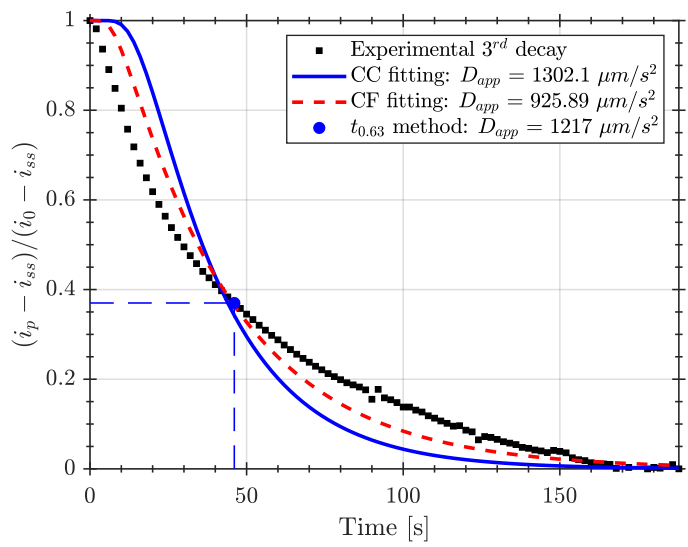

## A.2. For 1100ºC

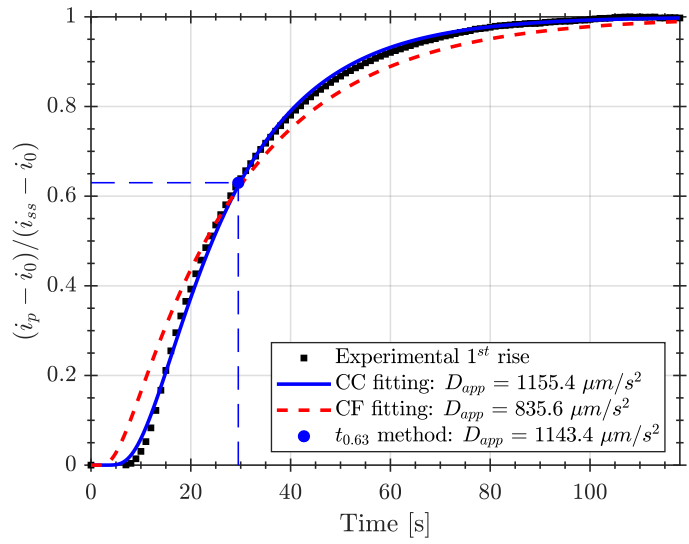

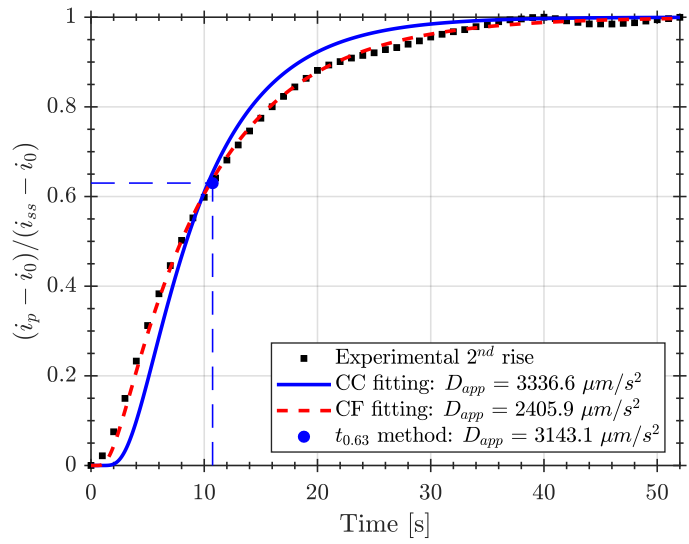

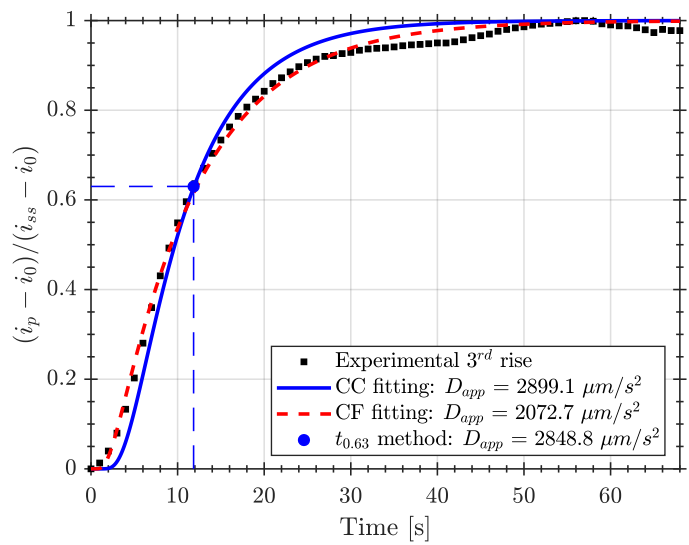

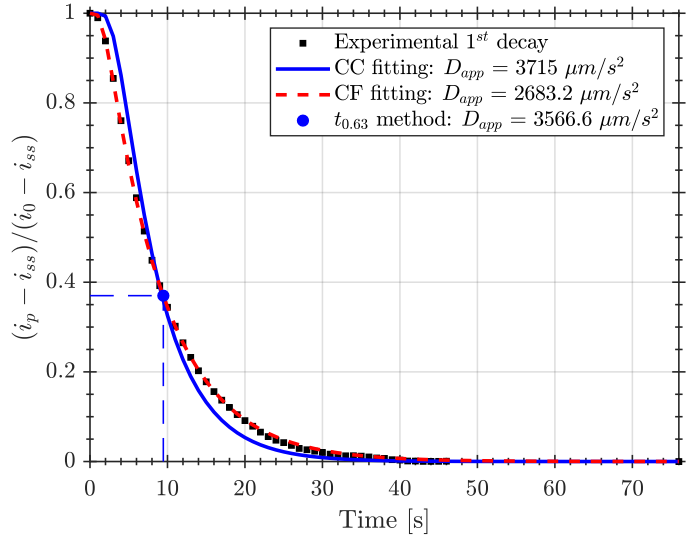
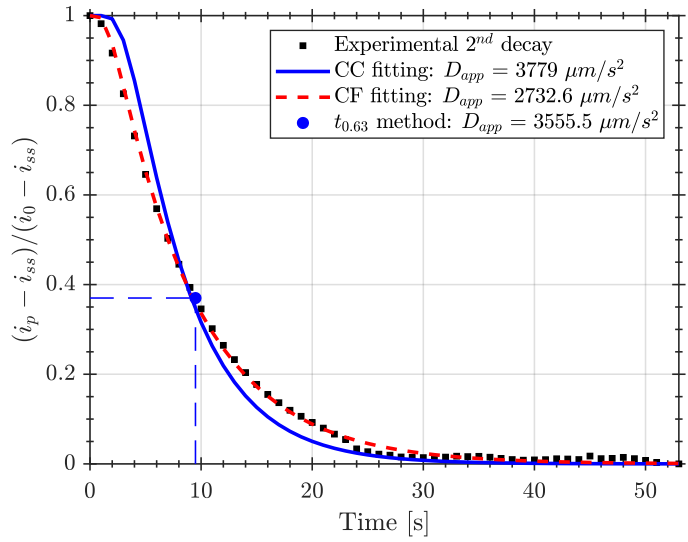
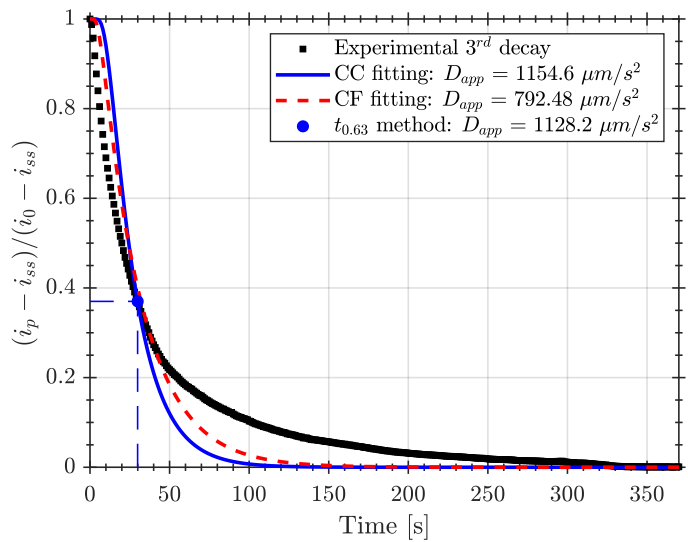